\documentclass[1p,12pt]{elsarticle}
\usepackage{graphics} 
\usepackage{epsfig} 
\usepackage{epstopdf}
\usepackage{color}
\usepackage{graphicx} 
\usepackage{epsf,wrapfig,epsfig}
\usepackage{subfigure}
\usepackage{amssymb,amsmath,amsfonts}
\usepackage{hyperref}
\usepackage{url} 
\usepackage{natbib}
\usepackage{fancyhdr}
\usepackage[T1]{fontenc}
\usepackage{ulem}
\usepackage{multirow}
\usepackage{longtable}
\usepackage{changepage}
\usepackage{tikz}
\usepackage{geometry}
\usepackage{pdflscape}
\definecolor{shadecolor}{rgb}{0.8, 0.8, 1}
\graphicspath{{./graphics}}
\begin{document}
\title{Modeling of the interaction of rigid wheels with dry granular media }
\author[1]{Shashank Agarwal}
\author[1]{Carmine Senatore}
\author[2]{Tingnan Zhang}
\author[2]{Mark Kingsbury}
\author[1]{Karl Iagnemma}
\author[2]{Daniel I. Goldman}
\author[1]{Ken Kamrin}
\ead{kkamrin@mit.edu}
\address[1]{Department of Mechanical Engineering, Massachusetts Institute of Technology, Cambridge, MA, USA}
\address[2]{School of Physics, Georgia Institute of Technology, Atlanta, GA, USA}

\begin{abstract}

We analyze the capabilities of various recently developed techniques, namely Resistive Force Theory (RFT) and continuum plasticity implemented with the Material Point Method (MPM), in capturing dynamics of wheel--dry granular media interactions.  We compare results to more conventionally accepted methods of modeling wheel locomotion. While RFT is an empirical force model for arbitrarily-shaped bodies moving through granular media, MPM-based continuum modeling allows the simulation of full granular flow and stress fields. RFT allows for rapid evaluation of interaction forces on arbitrary shaped intruders based on a local surface stress formulation depending on depth, orientation, and movement of surface elements. We perform forced-slip experiments for three different wheel types and three different granular materials, and results are compared with RFT, continuum modeling, and a traditional terramechanics semi-empirical method. Results show that for the range of inputs considered, RFT can be reliably used to predict rigid wheel granular media interactions with accuracy exceeding that of traditional terramechanics methodology in several circumstances. Results also indicate that plasticity-based continuum modeling provides an accurate tool for wheel-soil interaction while providing more information to study the physical processes giving rise to resistive stresses in granular media.
\end{abstract}

\maketitle
\section{Introduction}
In recent years, analysis of the interaction of lightweight robotic systems with natural terrain has raised skepticism as to whether the classical terramechanics theory is predictive for such systems. Basing his analysis on fundamental concepts of soil mechanics, Bekker ~\cite{bekker69} introduced a theory to predict mobility of wheeled and tracked vehicles in offroad scenarios. Bekker proposed a set of semi--empirical equations to predict various mobility aspects, such as compaction resistance, traction, sinkage, and driving torque. Over the past four decades, the original framework introduced by Bekker has been expanded and modified by several researchers, and has found applications in many studies of wheeled and tracked vehicles' mobility \cite{isminayo2007,ROB:ROB21483}. The most notable contribution to wheel--terrain modeling is the work by Wong and Reece which has become the de facto model of rigid cylindrical wheels on soft terrain \cite{wong67a,wong67b}. The model introduced by Wong and Reece derives wheel torque, thrust, and sinkage by estimating the stress distributions along the wheel-terrain contact region. The model is based upon the Bekker pressure-sinkage relation and the Janosi-Hanamoto shear-displacement equation \cite{janosi61}. 

In this paper, we explore the possibility of using two alternative modeling methodologies, namely granular resistive force theory (RFT) and continuum plasticity modeling using the Material Point Method (MPM), both of which have the potential to overcome many limitations of traditional semi-empirical methods. The RFT methodology was originally developed by Gray and Hancock~\cite{gray1955} for modeling swimming in viscous fluids, and was later extended by many ~\cite{maladen2009undulatory, zhang2014, li2013terradynamics} for evaluating resistive forces on a arbitrary shaped bodies moving through granular media. Granular RFT follows a different approach than traditional terramechanical models and assumes that the local force fields on each subsection of an intruder's leading surface are decoupled. Hence, the local stress functions on a surface element are extracted from independent penetration experiments at varying depths and  orientations. By linearly superimposing each element's stresses, RFT predicts the net resistive forces the granular volume applies to any arbitrary shape. Consequently, RFT can be applied to a variety of scenarios with different running gear geometry (potentially including complex grouser geometries), thus overcoming some of the limitations of traditional terramechanics methods.

Even though RFT is sufficiently accurate for a variety of problems (including rigid wheel locomotion scenarios as discussed in this paper), theoretical derivation of granular RFT from the basic laws of mechanics remains an open question. While the empirical nature of RFT creates advantages due to its rapid computation times over its existing mechanics based computational counterparts like the Discrete Element Method (DEM) (which captures many system states of insterests), it provides no direct information about the state of the media in which motion takes place. Hence, to better understand the mechanics of granular locomotion phenomena without having to use computationally expensive DEM, we perform a plasticity-based plane strain continuum modeling of wheeled locomotion scenarios using the MPM formulation.  More details about the method and implementation are provided in section 4 as well as in Dunatunga and Kamrin ~\cite{sachith2015mpm} whose MPM implementation is directly used here.

\section{Traditional Terramechanics Background}
Traditionally established terramechanics wheel models are based on the work of Bekker and Wong \cite{bekker1969introduction,wong01}. The underlying modeling approach relies on the analysis of two fundamental relations: the pressure--sinkage relation, and the shear stress--shear displacement relation. In the context of wheeled mobility, the pressure--sinkage relation (Eq.\ref{eq:pre_sink}) governs the depth that a wheel will sink into the terrain when subjected to load, and consequently how much resistance it will encounter while driving. The shear stress--shear displacement relationship (Eq.\ref{eq:shearstress_sheardisp}) governs the amount of traction that a wheel will generate when driven, and therefore how easily it will progress through terrain and surmount obstacles.
The pressure--sinkage relationship was described by Bekker in the form of a semi--empirical equation that relates sinkage with the normal pressure of a plate pushed into soil. The proposed relation is commonly referred to as the Bekker equation, and provides a link between the displacement (sinkage, $z$) and stress (pressure, $p$) of a plate (which can be viewed as a proxy for a wheel or track if one discretizes the leading surface of a wheel into sufficiently small sub--surfaces): 

\begin{equation}
p = \left( \frac{k_c}{b} + k_\phi \right) z^n 
\label{eq:pre_sink}
\end{equation}

Parameters $k_c$, $k_\phi$ and $n$ are empirical constants that are dependent on soil properties, and $b$ corresponds to the smaller dimension of the contact patch. These parameters can be obtained from field tests conducted with a device called a bevameter \cite{bekker69,wong01}. 

\begin{figure}[htbp]
\centering
\includegraphics[trim = 1mm 1mm 1mm 1mm, clip, width = 0.5\textwidth]{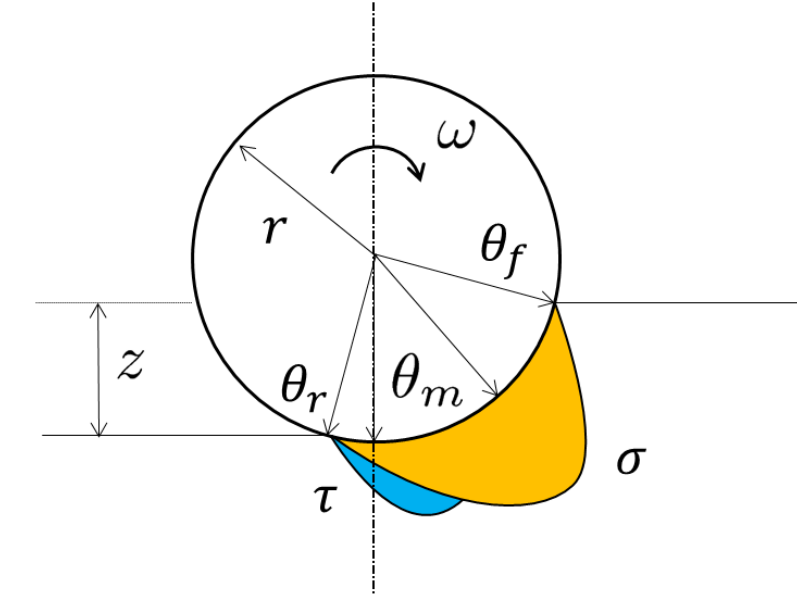}
\caption{Schematic representation of normal($\sigma$) and tangential($\tau$) stress profiles along a  wheel-soil interface.}
\label{fig:wheel_angle}
\end{figure}

The stress field under a wheel can be divided into two components (assuming a two dimensional model, temporarily ignoring out of plane motion): normal stress and tangential stress. A schematic representation of the stress distribution at a wheel-terrain interface is presented in Figure ~\ref{fig:wheel_angle}. 

Normal stress can be calculated by beginning with Bekker's pressure--sinkage relation, then introducing a scaling function to satisfy the zero--stress boundary conditions present at the fore and aft points of contact of the wheel with the terrain (known as `soil entry' and `soil exit'). The equation is expressed as a piecewise function, as:
\begin{eqnarray}
\sigma & = & \begin{cases} 
\sigma_1 = \left( \frac{k_c}{b}+k_{\phi} \right) z_1^n		&		\theta_m < \theta < \theta_f \\
\sigma_2 = \left( \frac{k_c}{b}+k_{\phi} \right) z_2^n		&		\theta_r < \theta < \theta_m \\
\end{cases} \nonumber \\
z_1 & = & r (\cos\theta-\cos\theta_f) \nonumber \\
z_2 & = & r \left( \cos \left( \theta_f -\frac{\theta-\theta_r}{\theta_m-\theta_r}(\theta_f-\theta_m ) \right) -\cos\theta_f \right)
\label{eq:sigma}
\end{eqnarray}

where $r$ is the radius of wheel, $\theta_f$ is the soil entry angle, $\theta_r$ is the exit angle, and $\theta_m$ is the angle at which the maximum normal stress occurs. This angle can be calculated as:

\begin{equation}
\theta_m = (c_1 + c_2 * \mathit{s} ) \theta_f
\end{equation}
where $c_1$ and $c_2$ are experimentally obtained constant parameters defined in~\cite{wong1967prediction}. $\mathit{s}$ represents the \textit{slip} and is defined as: \\
\begin{equation}
\mathit{s} = 1- \frac{V}{r\omega} = 1-\frac{V}{V_t} = \frac{V_t - V}{V_t} = \frac{V_j}{V_t}
\label{eq:jeq}
\end{equation}
 where, $V$ is the actual forward translational speed of the wheel, $V_t$  is the theoretical speed which can be determined from the angular speed  $\omega$ and the radius $r$  of the wheel, and $V_j$ is the speed of wheel-slip with reference to the ground. 

The shear stress in the longitudinal direction is the primary source of driving traction.  The shear stress $\tau$ is a function of $\sigma$, soil parameters and the measured shear displacement, $J$:
\begin{equation}
\tau = (c + \sigma \tan \phi) \left(1 - e ^{-\frac{J}{K}} \right)
\label{eq:shearstress_sheardisp}
\end{equation}
where  $c$  and $\phi$ are the cohesion and the angle of internal shearing resistance of the terrain, respectively, and $K$ is the shear displacement modulus which is a measure of the magnitude of the shear displacement required to develop the maximum shear stress (see ~\cite{wong10}). $J$ represents the shear displacement of the wheel edge with respect to the adjacent soil and is given as 
\begin{equation}
J = \int_0^{t_0} V_j \mathrm{d}t = \int_\theta^{\theta_f} V_j \frac{\mathrm{d} \theta}{\omega}
\end{equation}
where $V_j$ is the tangential slip velocity given earlier in equation \ref{eq:jeq}.

Thrust, $T$, is computed as the sum of all shear force components in the direction of forward wheel motion:
\begin{equation}
T = br \int_{\theta_r}^{\theta_f} \tau \cos \theta \mathrm{d} \theta
\end{equation}

Compaction resistance, $R_c$, is then computed as the result of all normal force components acting to resist forward motion:

\begin{equation}
R_c = br \int_{\theta_r}^{\theta_f}  \sigma \sin \theta \mathrm{d} \theta
\end{equation}

Drawbar pull, $F_x$, is calculated as the net longitudinal force (i.e. the difference between the thrust force and resistance force). $F_x$ is the resultant force that can either accelerate the wheel or provide a pulling force at the vehicle axle. 

\begin{equation}
F_x = T - R_c
\end{equation}

Driving torque can be obtained by integrating the shear stress along the wheel contact patch:

\begin{equation}
M = br^2 \int_{\theta_r}^{\theta_f}  \tau \mathrm{d} \theta
\end{equation}

This set of equations constitutes the backbone of the model proposed by Wong and Reece, and it will be referred from here on as the TM (i.e. the TerraMechanics) model.

\section{Resistive Force Theory Background}

While traditional terramechanics models study terrain within the framework of soil mechanics~\cite{bekker69}, in recent years, a new approach has been developed to study vehicle/robot locomotion by exploring the frictional fluid-like behavior emergent in sheared granular materials. A resistive force theory (RFT) was developed to characterize the interaction of arbitrary shapes with dry granular materials~\cite{li2013terradynamics,maladen2009undulatory}. 

Granular RFT was developed based on a formulation created for movement in low Reynolds number viscous fluids~\cite{gray1955} (where fluid inertia is negligible). For an object which locomotes by swimming through fluids (such that the velocity on each part of the swimming object takes different values), an analytical expression of the total drag forces is difficult to obtain from the Navier-Stokes equations. Gray and Hancock ~\cite{gray1955} approximated a solution to this problem by postulating that the force field on an infinitesimal element of a slender body (whose radius of curvature is significantly larger than the width) is hydrodynamically decoupled from the rest of its body.  The drag force on an element (of simple geometry) $dS$ is then computed from its local velocity and the tangent direction $\mathbf{\hat{t}}$ (or normal $\mathbf{\hat{n}}$) of the element. The net drag for the swimmer is then given by a linear superposition:

\begin{equation}
\mathbf{F}_d = \int [d\mathbf{F}_\parallel + d\mathbf{F}_\perp]\mathrm{d}S = \int [f_\parallel(\mathbf{v \cdot \hat{t}})\mathbf{\hat{t}} + f_\perp(\mathbf{v \cdot \hat{n}})\mathbf{\hat{n}}] \mathrm{d}S
\end{equation}

\begin{figure}[htbp]
\centering
\includegraphics[trim = 1mm 1mm 1mm 1mm, clip, width = 0.8\textwidth]{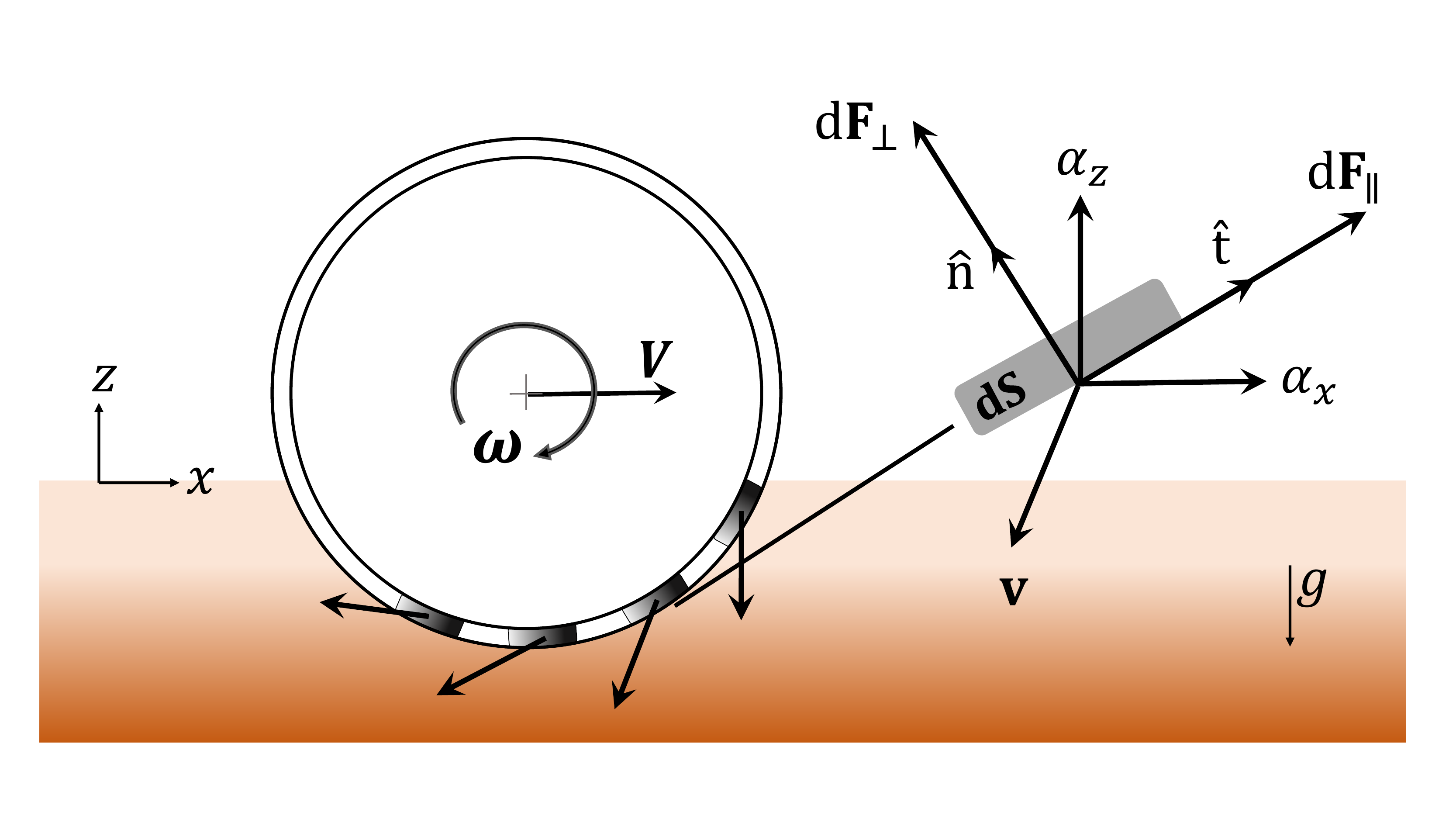}
\caption{\label{fig:figRFT} RFT illustration of a wheel moving on a granular medium. $\mathbf{V}$ is the forward translational speed of the wheel center, $\omega$ the angular velocity. Each segment $dS$ at the wheel surface has different velocity $\mathbf{v}$, and orientation (denoted by its normal $\mathbf{\hat{n}}$ or tangential $\mathbf{\hat{t}}$). $d\mathbf{F}_{\perp,\parallel}$ represent the local forces.}
\end{figure}

The formulation was recently adapted to subsurface swimming in granular media by Maladen et al. \cite{maladen2009undulatory}. Unlike viscous fluids, granular RFT is not restricted to slender bodies and for an intruder moving slowly ($v \lesssim 0.5$ m/s) in a granular media, the drag force is dominated by friction i.e. it is insensitive to the moving speed, and increases with penetration depth and compaction. The RFT formula then takes the form:

\begin{equation}
\mathbf{F}_d = \int [d\mathbf{F}_\parallel + d\mathbf{F}_\perp]\mathrm{d}S = \int [\alpha_x(\mathbf{\hat{v},\hat{t}})\mathbf{\hat{x}} + \alpha_z(\mathbf{\hat{v},\hat{t}})\mathbf{\hat{z}}]  |z|\mathrm{d}S\,,
\end{equation}
where $\alpha_x$ and $\alpha_z$ are local stresses per unit depth on a small surface element $dS$ at the depth of $|z|$. When granular RFT was first developed, the functional forms of $\alpha_x$ and $\alpha_z$ were determined from experimental trials ~\cite{maladen2009undulatory, li2013terradynamics}. Askari and Kamrin ~\cite{askari2015intrusion} later successfully verified that the experiment based functional form proposed earlier actually matches with the functional form obtained using a tension-free Druker-Prager plasticity model (described in the next section), thereby indicating a possibility of the use of plasticity based modeling in the scenarios where RFT is applicable. Hence for clarification, plasticity based continuum simulations are explored and explained in more detail next.

\section{Continuum Modeling using the Material Point Method (MPM)} \label{MPM_intro}
In recent years successful attempts have been made by various authors ~\cite{sachith2015mpm,anderson2009,Wieckowski2011} in using the Material Point Method (MPM) to implement continuum models of granular flows. MPM is a derivative of the fluid-implicit-particle (FLIP) method~\cite{flip}, which is based on the particle-in-cell (PIC) method ~\cite{pic}. The key idea behind MPM is that the state of the simulated material is contained in Lagrangian material points, while the equations of motion are solved on a background computational mesh in a manner similar to finite element methods. Since the state is saved at each material point, the mesh is reset at the beginning of each computational step,  allowing for large deformations without mesh distortion. The basic computational layout is extensively discussed in Sulsky et al.~\cite{mpmsulsky}. The model developed for dry non-cohesive granular media by Dunatunga and Kamrin \cite{sachith2015mpm} is used in this work. The model is obtained by assuming a Drucker-Prager yield criterion, incompressible plastic shear flow, and cohesionless response in extension whereby the material becomes stress free when below a critical density: 

\begin{equation}
\bar{\tau} \leq \mu_sP   \quad \textrm{and} \quad \rho =\rho_c  \quad   \textrm{if} \quad P>0   
\end{equation}
\begin{equation}
P, \bar{\tau} = 0  \quad \textrm{if} \quad \rho<\rho_c  
\end{equation}
where:\\
\hspace*{10mm} $\boldsymbol{\sigma'}$= $\boldsymbol{\sigma}$ + $P\mathbf{1}$  $\quad$ $\quad$ is the deviatoric part of the stress tensor \\
\hspace*{10mm} $P = -1/3$ tr($\boldsymbol{\sigma}$)      		 $\quad$\ is the hydrostatic pressure \\
\hspace*{10mm} $\bar{\tau}$  =  $|\boldsymbol{\sigma'}/\sqrt{2}|$   $\quad$ $\quad$   \hspace{1mm}is the equivalent shear stress  \\
\hspace*{10mm} $\rho_c $ \hspace{30mm} is the critical close-packed granular density\\

The system above is implemented in the approximately rigid-plastic limit by treating it as the plastic part of an elasto-plastic model with very stiff elastic response, as in \cite{sachith2015mpm}.

\section{Discussion of traditional Terramechanics and use of Continuum and RFT models in locomotion modeling}

Traditional terramechanics approaches rely on a set of parameters that include intrinsic soil properties such as cohesion and internal angle of friction, along with semi--empirical parameters including the shear displacement modulus and sinkage coefficients. Resulting models are not computationally intensive, but are often over--parametrized and require ad--hoc terrain testing(eg: the Bekker model, which assumes wheels to be rigid cylinders, requires $\sim$10 fitting parameters to be evaluated using a specialized instrument called a Bevameter). This typically results in restricted applicability of the aforementioned models when wheel geometries are modified or operational conditions diverge from nominal conditions (e.g., the high slip condition), and when parameter estimation from wheel performance data is attempted.  On the other hand, approaches based on RFT have the advantage of relying on a compact set of parameters, and can be applied to a wide range of wheel geometries. Terramechanics models can be utilized for broader terrain types given proper characterization; the applicability of RFT to cohesive soils has not been verified yet. Both approaches are currently limited to homogeneous soils.

The basic limitation of both of these empirical methods is that they are limited to finding the forces on the locomoting bodies and give no detailed information about the surrounding granular media deformation. Such limitations are easily overcome by using correctly applied continuum modeling, which not only gives the forces acting on the body, but also the other time dependent variables like stress, strain, and velocity profiles in the media provided accurate constitutive relations are used. Continuum modeling can also take into account the elasticity of wheels (if needed), which are usually considered to be rigid in both RFT and terramechanics models in this study.\\

\section{Experimental Setup}
\subsection{Hardware}

\begin{figure}[htbp]
\centering
\includegraphics[trim = 70mm 55mm 70mm 70mm, clip, width = 0.95\textwidth]{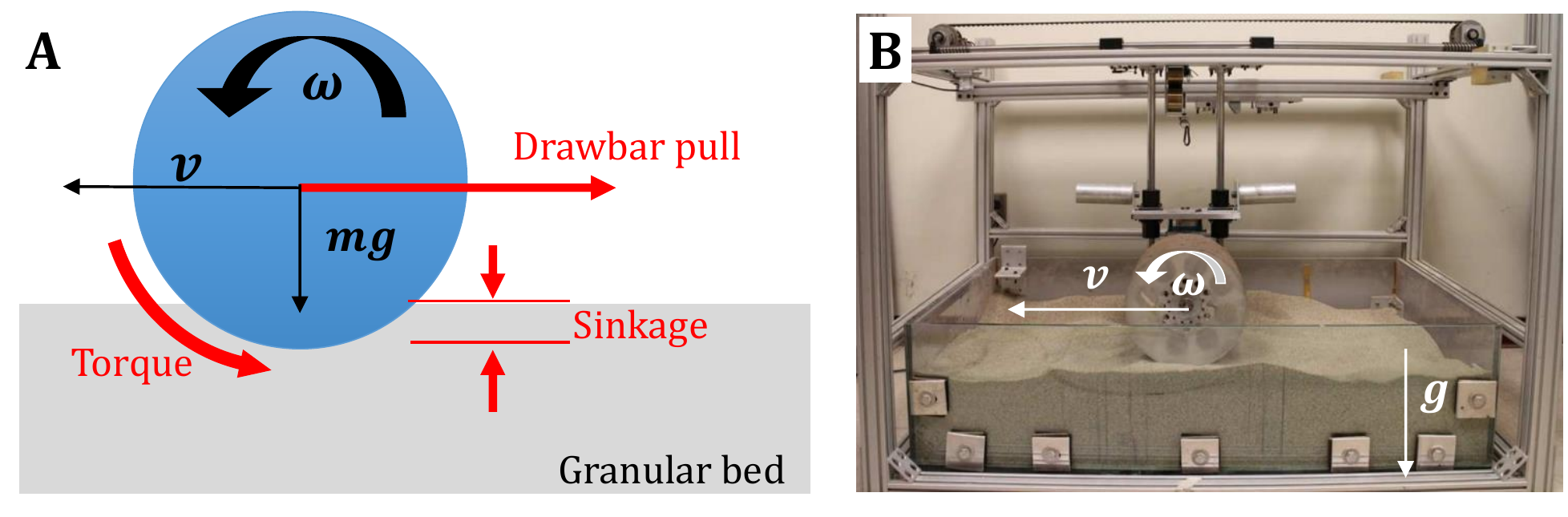}
\caption{\label{fig:MIT_rig}
A) Schematic and B) experimental setup of one of the forced--slip terramechanics rigs used in the study, where translation ($v$) and angular velocities ($\omega$) are controlled while torque, drawbar pull and sinkage ($z$--direction motion is free) are measured.
}
\end{figure}

A multipurpose terramechanics rig based on the design described by Iagnemma et al.~\cite{iagnemma2005laboratory} was designed and fabricated for conducting the experiments in this study. The testbed is pictured in Figure~\ref{fig:MIT_rig} and is composed of a Lexan soil container surrounded by an aluminum frame to which all the moving parts, actuators, and sensors are attached. A carriage slides on two low-friction rails to allow longitudinal translation while the wheel, attached to the carriage, is able to rotate at a desired angular velocity. The wheel mount is also able to freely translate in the vertical direction. This setup allows control of slip and vertical load by modifying the translational velocity of the carriage, angular velocity of the wheel, and applied vertical load. Horizontal carriage displacement is controlled by a timing belt actuated by a 90 W Maxon DC motor, while the wheel is directly driven by a 200 W Maxon DC motor. The motors are controlled through two identical Maxon ADS 50/10 4-Q-DC servoamplifiers. The carriage's horizontal displacement is monitored with a Micro Epsilon WPS-1250-MK46 draw wire encoder, while wheel vertical displacement (i.e., sinkage) is measured with a Turck A50 draw wire encoder. An ATI Omega 85 6-axis force torque transducer is mounted between the wheel mount and the carriage in order to measure vertical load and traction generated by the wheel. Finally, a flange-to-flange reaction torque sensor from Futek (TFF500) is used to measure the driving torque applied to the wheel. Control and measurement signals are handled by a NI PCIe-6363 card through Labview software.

The apparatus described above is capable of approximately $1$ m of total horizontal displacement at a maximum velocity of approximately $120$ mm/s, with a maximal wheel angular velocity of approximately $40^o$/s. The container width is $0.6$ m, while the soil depth is $0.16$ m. 

The Goldman group has previously designed and fabricated several fluidizing testbeds that allow control of the packing state of granular materials and have used these extensively in locomotion studies ~\cite{locomotionbed1,locomotionbed2,locomotionbed3}.
For the poppy seed experiments presented in this paper, the multipurpose terramechanics rig was assembled over a $2.5$ m long, $0.5$ m wide fluidized bed trackway filled with poppy seeds. Poppy seeds have certain properties similar to natural sand \cite{li2013terradynamics}, and have a low enough density ($\approx 1.0$ kg/cm$^3$) to be fluidized easily with low-cost blowers. The trackway has a flow distributor of porous plastic (Porex, thickness $0.64$cm, average pore size of $90 \mu$m) through which four $300$ LPM leaf blowers (Toro) blow air. When the leaf blowers are at maximum power, the poppy seeds are fluidized into the bubbling regime. As the power from the leaf blowers is slowly reduced to zero, the granular media settles into a loosely packed state (volume fraction $\phi \approx 0.580$). Additionally, the power can be reduced to just below the onset of the bubbling regime and a motor with an off-center mass attached to the bed can be turned on to compact the granular media down to its critical packing state (volume fraction $\phi \approx 0.605$). Once the desired packing state is achieved, the airflow is turned off for the duration of the experiment.

The experiments were conducted under forced--slip conditions, such that the wheel angular velocity $\omega$ and wheel longitudinal velocity $V$ were controlled according to: 
\begin{equation}
s = 1 - \frac{V}{\omega r}
\end{equation}
where $s$ is the desired slip ratio and $r$ is the nominal wheel radius. Wheel angular velocity was held constant while longitudinal velocity was varied to achieve the desired slip ratio. Experiments were conducted under vertical loads varying between $18$ N and $190$ N (see Table \ref{tab:wheels_properties}).

\subsection{Simulants} \label{simulants}

Three simulants were used in this research: Quikrete medium sand (MS), Mars Mojave Simulant (MMS), and poppy seeds (PS). MS is a commercially available product called Quikrete 1962 Medium Sand. It is a silica sand with predominant size in the $0.3 - 0.8$ mm range. MMS is a mixture of finely crushed and sorted granular basalt intended to mimic, both at chemical and mechanical levels, Mars soil characteristics \cite{mojave2007}. The MMS particle size distribution spans from the micron to millimeter scale, with $80 \%$ of particles above $10\mu$m.

\begin{table}[ht!]
\caption{Mechanical properties of the granular materials considered in this study. Quikrete Medium Sand (MS) and Mars Mojave Simulant (MMS) were characterized through plate penetration tests and direct shear tests. The RFT constant for these simulants was extrapolated from the plate penetration tests. The Poppy Seeds (PS) on the other hand, were only characterized with plate intrusion experiments. The values for $\mu_{internal}$ for plane strain MPM simulations were obtained using sinkage matching with zero--slip experiments for each material. }
\centering
\vspace{0.5em}
\begin{tabular}{r c c c l}

   & MS & MMS & LPS & CPS \\
  \hline
  $k_c$ [kN/m$^{n+1}$] & -2.05e+4 & 846 & -2.06e+5 & -3.24e+5 \\
  \hline
  $k_\phi$ [kN/m$^{n+2}$] & 3.13e+6 & 6708 & 7.07e+6 & 1.11e+7 \\
  \hline
  $n$ & 1.0 & 1.4 & 1 & 1 \\
  \hline
  $c$ [Pa] & 1500 & 600 & 0 & 0\\
  \hline
  $\Phi$ [deg] & 34 & 35 & 36 & 45\\
  \hline
  $K$ [m] & 0.0006 & 0.0006 & 0.045 & 0.045\\
  \hline
  RFT Constant [$\mathrm{N/cm^3}$] & 2.02& 3.05 & 0.35 & 0.55\\
  \hline
  $\rho_{grain}$ $[kg/m^3]$ & 2600 &2875 & 1100 & 1100\\
  \hline
  Packing Fraction $\phi$ & 0.6 & 0.6 & 0.580 & 0.605\\
  \hline  
  MPM : $\mu_{internal}$ & 0.53 & 0.50 & 0.53 & 0.54\\
  \hline  
\end{tabular}
\label{tab:mech_properties}
\end{table}

\begin{figure}[ht!]
\centering
\includegraphics[width = 1.0\textwidth]{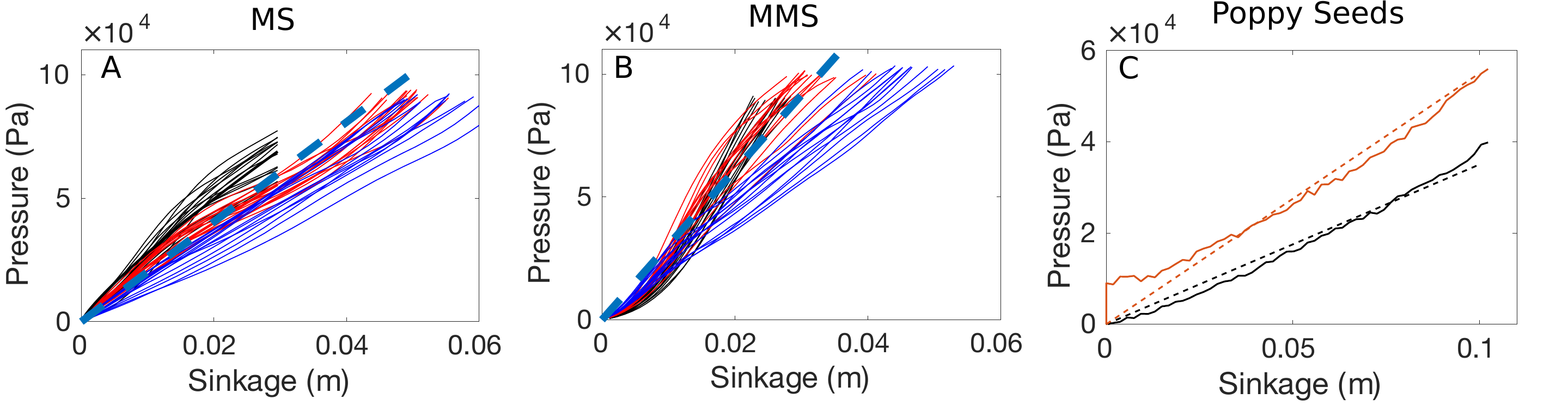}
\caption{\label{fig:pressure_sinkage} Pressure--sinkage relations for the granular media used in the study. Penetration pressure in Quikrete Medium Sand (MS, A) and Mars Mojave Simulant (MMS, B) using plates of $16$cm by $\{3, 5, 7\}$cm in area (blue, red and black lines, resp.) are plotted. Blue dashed lines (in A and B) represent linear regression results from red and black curves. (C) Penetration pressure for a $3.81 \times 2.54\, \mathrm{cm^2}$ plate moving vertically in loosely (solid black curve) and closely (solid orange curve) packed poppy seeds. Dashed lines of the same color are linear fitting results.} \label{fig:pressure-sinkage}
\end{figure}

Soil properties for the MS and the MMS were measured through a series of plate penetration tests and direct shear tests: nominal soil parameters are presented in Table~\ref{tab:mech_properties}. Plate penetration experiments were conducted with rectangular plates measuring 0.16 m by $\lbrack 0.03, 0.05, 0.07 \rbrack$ m (Figure~\ref{fig:pressure-sinkage}A and ~\ref{fig:pressure-sinkage}B). These particular plate dimensions, according to terramechanics principles, are adequate for estimating terrain pressure--sinkage parameters for modeling a wheel with an approximate contact patch area of 0.16 m by 0.05 m. Direct shear experiments were conducted following standard terramechanics procedures~\cite{wong10} and using a $6.0 \times 6.0$ $\rm{ cm^2}$ shear box. The shear displacement modulus ($K$) calculated from direct shear tests is on the order of tenths of millimeters  \cite{senaiagn2011,ROB:ROB21483}, while typical terramechanics literature values range between 10 and 30 millimeters \cite{wong01}. This discrepancy is likely due to the fact that the boundary conditions that develop under a running wheel are different from the ones that develop in a shear box. For wheel--terrain interaction studies, vane or ring shear devices are advised for shear strength characterization. This testing approach usually produces larger values of shear modulus, which results in more accurate predictions with the TM models. For poppy seeds, the terramechanics parameters had to be extrapolated from experiments conducted with one plate having a size of $2.5 \times 3.8 $ $\rm{ cm^2}$ (Figure \ref{fig:pressure-sinkage}C). Consequently, pressure-sinkage parameters $k_c$, $k_{\phi}$, $n$ were calculated imposing linear response (i.e. $n = 1$). The PS were not characterized for shear loading (at least not in the terramechanics sense), hence cohesion was set to zero, angle of internal friction was assumed to match the angle of repose, and the shear modulus was used as a free parameter for TM modeling on PS.\\

\begin{figure}[ht!]
\centering
\includegraphics[trim = 75mm 0mm 60mm 20mm, clip, width = 0.8\textwidth]{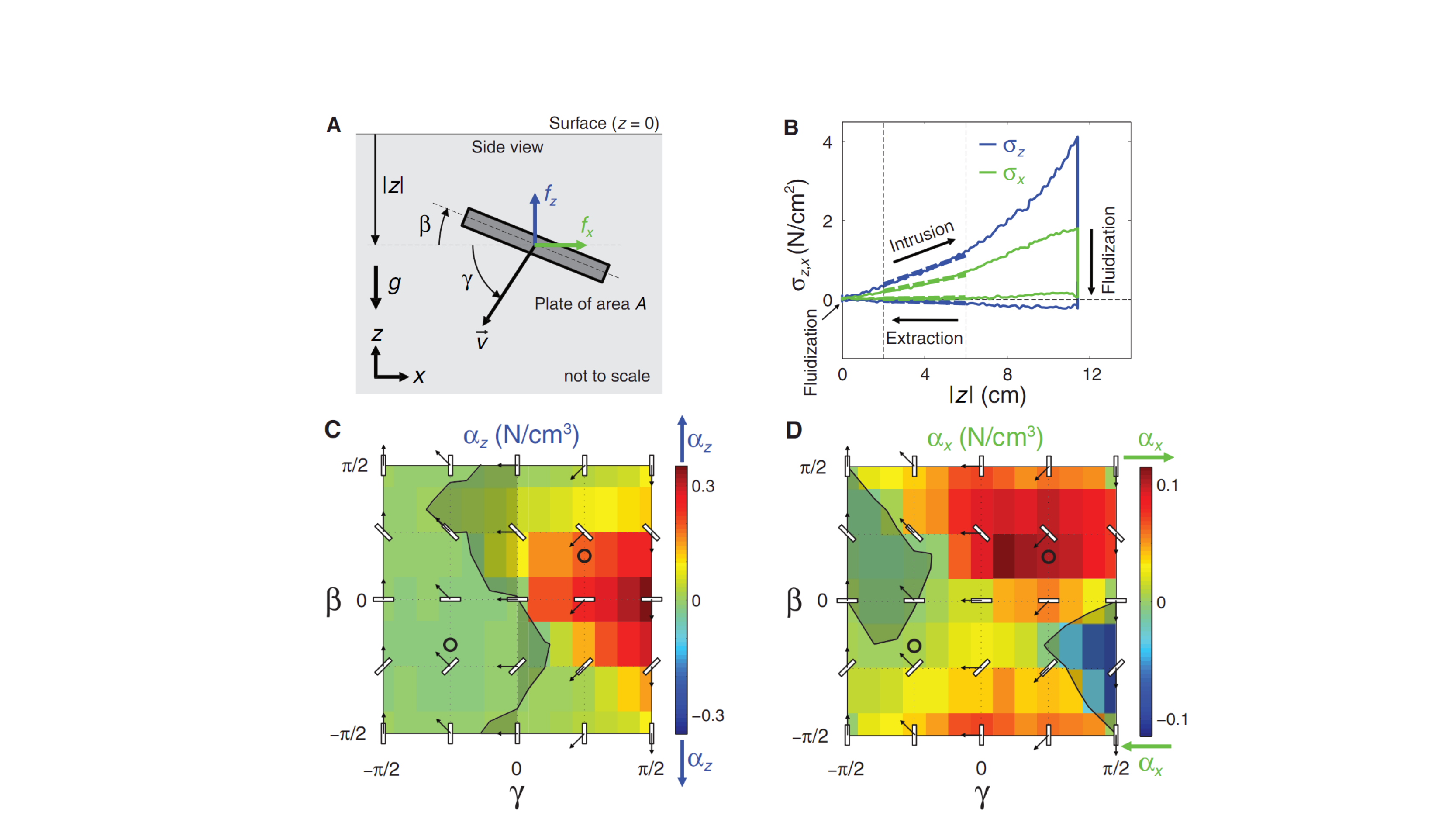}
\caption{RFT characterization of poppy seeds using measurement of resistive forces on a plate element. (A) Lift ($f_z$) and drag ($f_x$) forces on a model element surface (a rigid plate of area $A$) moving in the $x-z$ vertical plane at velocity $\mathbf{{v}}$. $\beta$ represents the orientation angle and $\gamma$ the angle of attack. $\vert z\vert$ is the depth at the center of the plate. $\mathbf{g}$ is gravitational acceleration. (B) Measured vertical and horizontal stresses $\sigma_{x,z} = f_{x,z} / A$ for the plate moving in a container of loosely packed ($\phi=0.58$) PS. Between each intrusion and extraction, the granular medium was fluidized and settled to restore the undisturbed condition. (C,D) The response $\alpha_{z,x} = \sigma_{z,x} / \vert z\vert$ versus $\beta$ and $\gamma$. Figure adapted from~\cite{li2013terradynamics}. 
\label{fig:rft_plot} 
}
\end{figure}

For the experiments with poppy seeds (PS), fluidized testbeds were used (Figure~\ref{fig:rft_plot}A) so that the PS packing fraction could be controlled. In a previous study of legged robot locomotion performance on granular media~\cite{li2013terradynamics}, RFT force relations for this material were characterized under both loosely and closely packed conditions. Instead of using $dF_{\perp,\parallel}$, for convenience we used the lab $x-z$ coordinate frame for all force measurements and calculations (Figure~\ref{fig:rft_plot}B). The stresses $\sigma_{x,z}(\beta, \gamma) = f_{x,z}(\beta, \gamma)/A$ on a small plate (as a model surface element) were measured in independent drag experiments of different combinations of the orientation angle $\beta$ and the attack angle $\gamma$. For the sinkage range relevant to our wheel experiments ($\lesssim 80$ mm), $\sigma_{x,z}$ increased approximately linearly with penetration depth (Figure~\ref{fig:rft_plot} B); thus we extracted the response surfaces as $\alpha_{x,z}(\beta, \gamma) = \sigma_{x,z}/\vert z\vert$ (Figure~\ref{fig:rft_plot} D). The RFT constant, defined as $\alpha_{z}(0, \pi/2)$, is listed in Table~\ref{tab:mech_properties}.

We did not thoroughly test the angular dependencies of $\alpha_{x,z}$ for the MS and MMS sands. Instead, we assumed, as in \cite{li2013terradynamics}, that the responses of the MS and MMS had a similar angular dependence to the poppy seeds. The RFT constant for each material was characterized from its pressure-sinkage relation (Figure~\ref{fig:pressure-sinkage} A and B). The RFT constants were obtained to ensure that the linearly scaled $\alpha_{x,z}$ for MS/MMS (with the use of their  respective RFT constant) gives the same pressure-sinkage relations as obtained from the experiments Figure ~\ref{fig:pressure-sinkage}. We also excluded data points corresponding to the $16 \, \mathrm{cm} \times 3 \, \mathrm{cm}$ plate when applying the linear regression model to the pressure-sinkage curves because the plate width below 5 cm would be representative of extremely narrow contact patch areas (which we did not observe with wheels A and B). The RFT constants for the MS and MMS, obtained from the slopes of the fitted pressure-sinkage curves (Figure~\ref{fig:pressure-sinkage}A and B, dashed lines), are $\sim6-9$ times greater than that of the loosely packed PS. 

For the MPM based continuum modeling, we assumed that the motion of all the wheels considered in this study could be modeled as plane strain problems (which is a justifiable assumption to take if the out--of--plane depth of the contact area between the wheel and sand is larger than its width). The plastic flow parameters for the simulations were calibrated by matching zero-slip experimental data to zero-slip plane-strain MPM simulations. Since the actual deformation in experiments was not always plane-strain, we accept potential inaccuracy brought about by the plane-strain simplifying assumption.  The MPM simulations were found to be most sensitive to internal coefficient of friction. The effective internal friction values ($\mu_{internal}$) for each material were evaluated by finding the value of $\mu_{internal}$ which when used in the MPM simulation results in the same sinkage found experimentally. This matching was done once (for the zero-slip case) for each media and these values were then used for all simulations in this study. Values for all four materials are shown in Table \ref{tab:mech_properties}. Calibration trials for deciding the surface friction coefficient ($\mu_{surface}$) between the wheels and the grains were found to be accurate with the use of experimentally obtained surface friction coefficients and hence different experimentally found wheel--sand pair values (reported in section \ref{section:wheels}) were used.

\subsection{Wheels} \label{section:wheels}

\begin{figure}[htbp]
\centering
\includegraphics[trim = 1mm 1mm 1mm 1mm, clip, width = 0.75\textwidth]{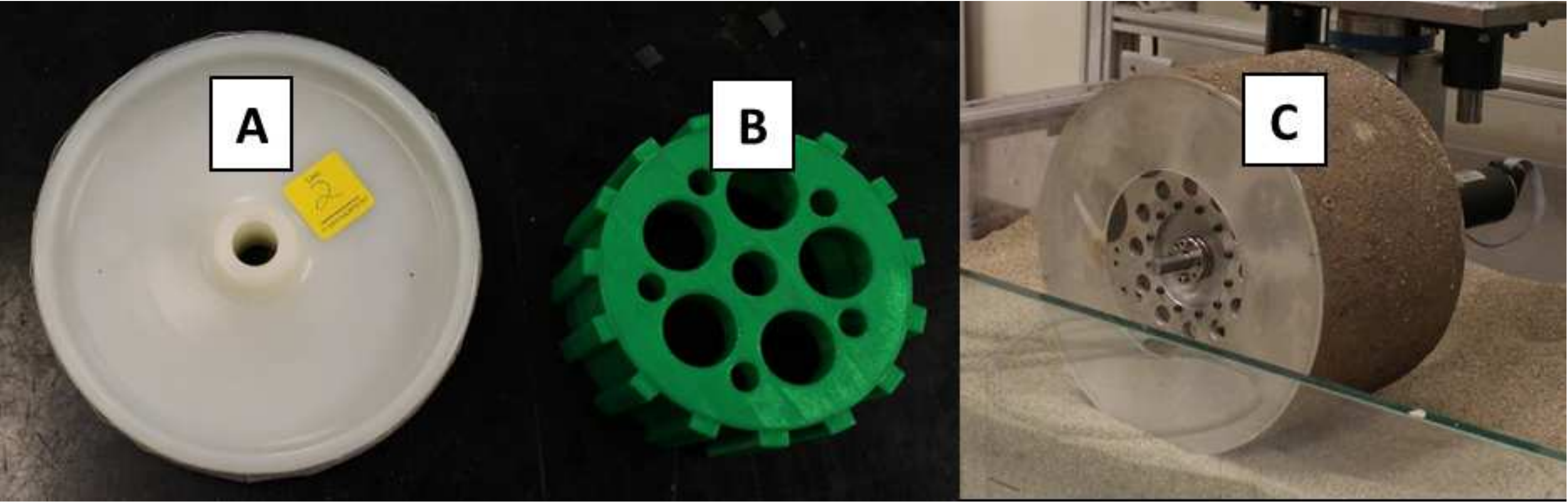}
\caption{\label{fig:wheels} Wheels utilized in this study (Images not to scale).}
\end{figure}

Experiments were conducted with three different wheels with aspect ratios (width/radius) of $0.5$, $1.05$ and $1.23$. The wheels are shown in Figure \ref{fig:wheels}, while wheel dimensions are given in Table \ref{tab:wheels_properties}. Wheels A, B, and C were tested on PS, while wheel C was also tested on MS and MMS. 

\begin{table}[h]
\caption{Specifications of the wheels utilized in this study and summary of experiments conducted. Surface coating, 60 grit, PLA, MMS.}
\centering
\vspace{0.5em}

\begin{tabular}{r c c c}
  \hline
   & A & B & C\\
  \hline
  Type & Smooth Wheel & Lugged Wheel & Smooth Wheel\\
  \hline
  Radius [mm] & 101.6 & 72.5 (to lug tips) & 130\\
  \hline
  Aspect Ratio & 0.5 & 1.05 & 1.23\\
  \hline
  PS  & \checkmark & \checkmark & \checkmark\\
  \hline
  MS  & - & - & \checkmark\\
  \hline
  MMS & - & - & \checkmark\\
  \hline
  Vertical Loads [N] & 20 & 18 & 80-190\\
  \hline
\end{tabular}
\label{tab:wheels_properties}
\end{table} 

Wheel A is a Nylon wheel with a narrow aspect ratio. The wheel surface was coated with 60 grit sand paper in order to guarantee sufficient friction at the wheel-terrain interface. Wheel B was manufactured using a MakerBot Replicator II 3D printer using PLA filament. The wheel has 15 lugs, equally spaced, $10$ mm tall and $11$ mm thick, which span the whole width of the wheel. This wheel has no sandpaper coating, as the presence of the lugs guarantees sufficient wheel-terrain engagement. Finally, wheel C is an aluminum cylinder coated with MMS. For continuum modeling of wheel--media surface interaction,  the coefficient of surface friction for wheel C with all the simulants was taken as 0.55 and for wheel A and B (which were experimented only with PS), the values were 0.60 and 0.35 respectively.

\section{RFT Simulations}

\begin{figure}[htbp]
\centering
\includegraphics[trim = 55mm 40mm 55mm 30mm, clip, width = 0.9\textwidth]{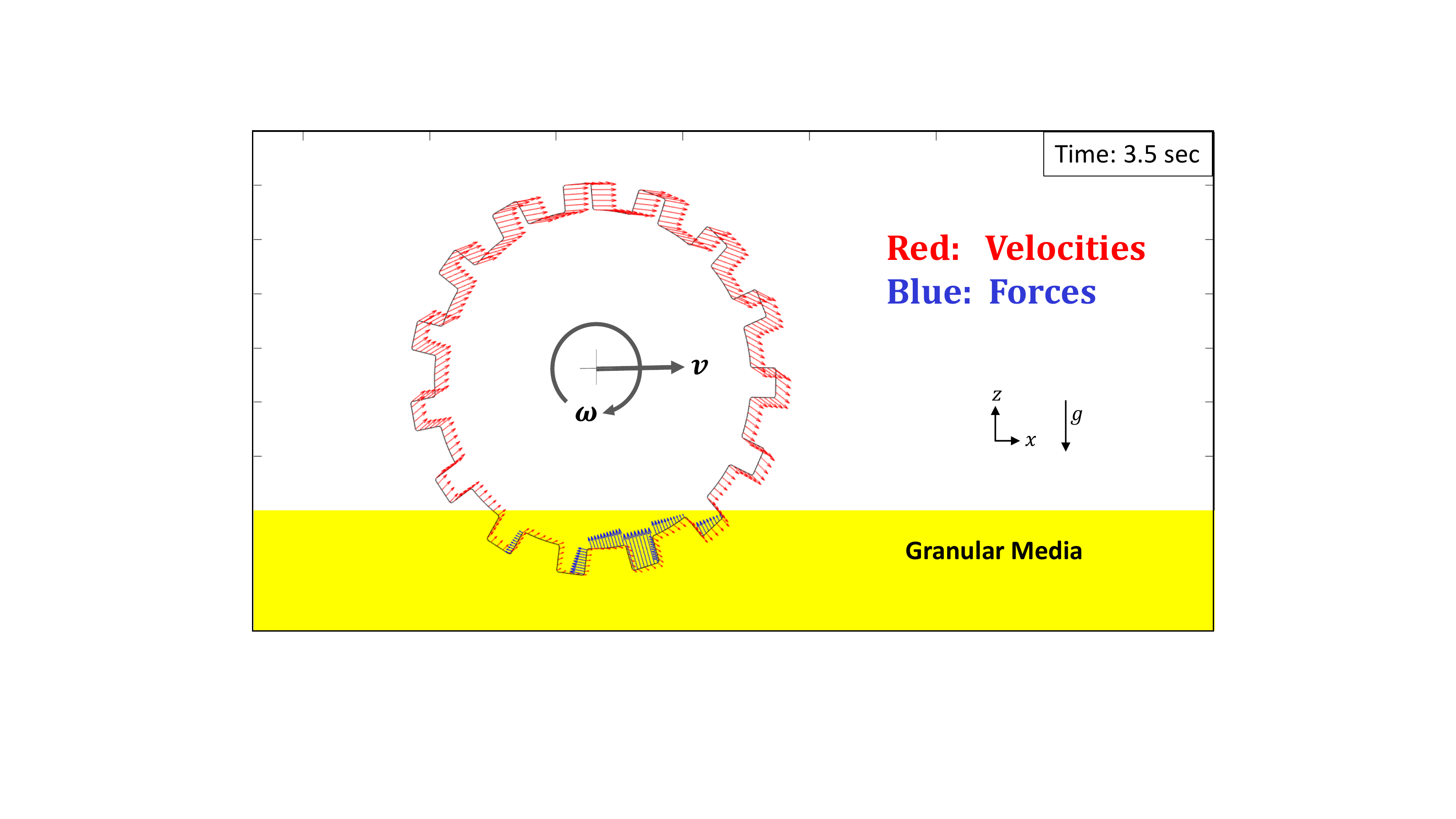}
\caption{A sample implicit RFT implementation (wheel Type B) in MATLAB where red arrows represent the normalized velocity vectors  of the wheel surface elements, and blue arrows show the normalized resistive force vectors on each subsection.}
\label{fig:rft_sample}
\end{figure}

RFT simulations were implemented using an implicit iterative scheme in MATLAB. Utilizing the rigid wheel assumption, wheel surfaces were discretized into smaller subsurfaces that together approximated the total geometry. The orientation, velocity direction, depth, and area of each sub-surface along with normalised force per unit depth from Li et al \cite{li2013terradynamics} and associated scaling coefficients from Table~\ref{tab:mech_properties} were used for finding the resistive forces from the media on each subsurface. The net resistive force and moment on the wheel were calculated using the RFT superposition principle mentioned previously. As the wheel's $x$--translational motion was predefined (forced slip tests), a momentum balance in the $x$ lab frame coordinate and angular momentum balance along the axis of the wheel, gave the values of total drawbar pull and torque (respectively) required to sustain the given velocity conditions. The vertical motion (sinkage) of the wheel was captured by balancing momentum in the lab frame $z$ coordinate. In performing all these simulations, a `leading edge hypothesis' was also used which made sure that the resistive forces experienced by the wheel consisted of contributions from only those surface elements which were moving `into' the sand, i.e. surfaces whose outward normal and velocity make a positive dot product. A sample RFT simulation setup for wheel type B is shown in Figure ~\ref{fig:rft_sample}. 

\section{MPM Simulations}
\begin{figure}[htbp]
\centering
\includegraphics[trim = 20mm 1mm 60mm 1mm, clip, width = 1.0\textwidth]{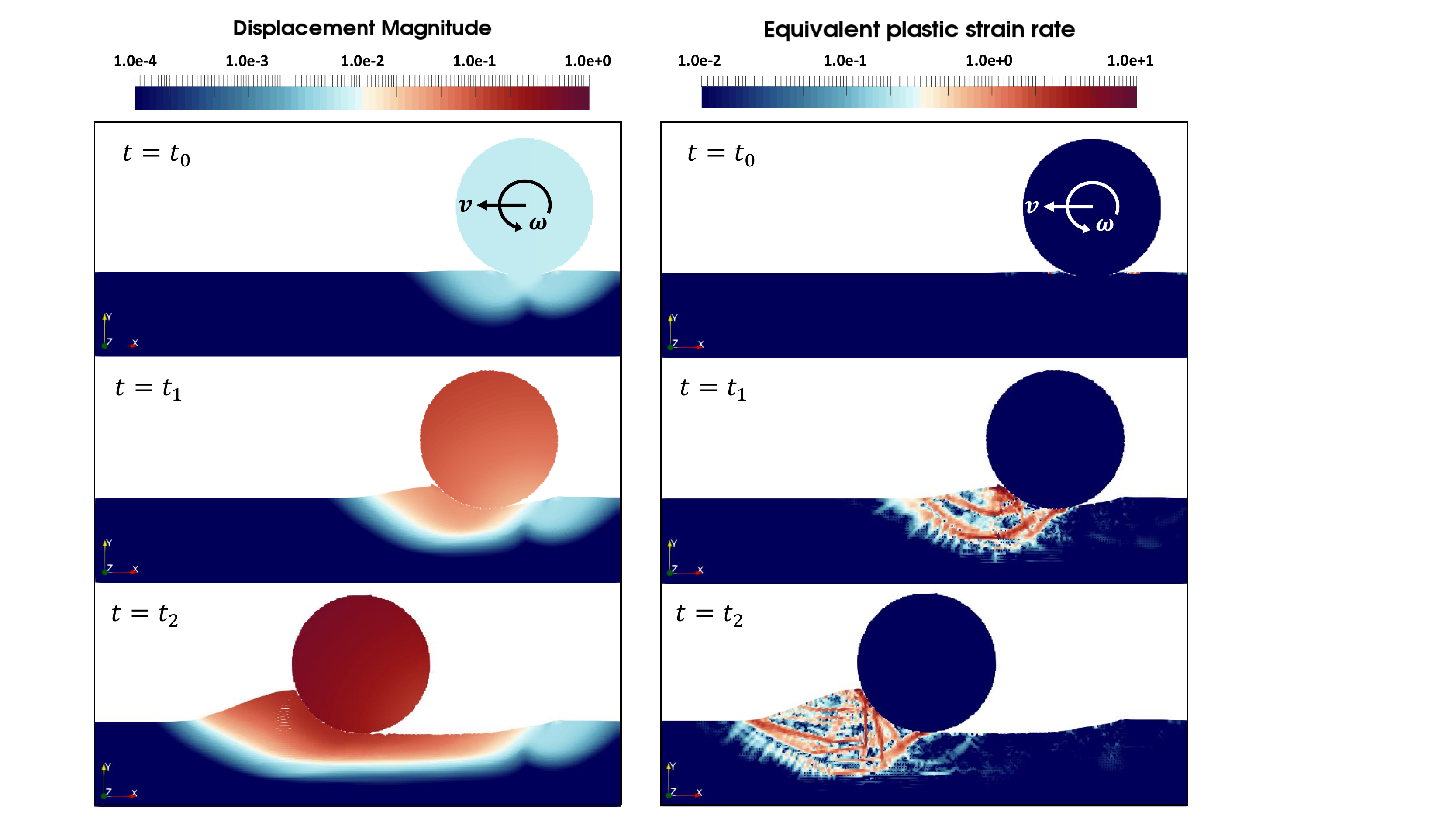}
\caption{
A sample MPM implementation of wheel Type C in LPS at negative slip ( $i= -0.3$) velocity condition. Time $t_0$ corresponds to state when the wheel is freely resting on the medium, $t_1$ corresponds to the transition state, and $t_2$ corresponds to a time instance when the equilibrium sinkage condition is met. 
}
\label{fig:mpm_sample}
\end{figure}

The MPM algorithm described in Dunatunga and Kamrin \cite{sachith2015mpm} was used to implement the set of constitutive equations given in section \ref{MPM_intro}. The values of relevant material properties for various simulants used in this study are provided in Table \ref{tab:mech_properties}. The wheel was modeled as a stiff elastic solid with fixed horizontal translation speed and a fixed angular velocity, which are instantaneously applied on the wheel explicitly. In terms of simulation resolution, a $200\times200$ grid was used to represent a domain size of 1m$\times$1m with $2\times2$ linear material points seeded per grid cell at the beginning of the simulation. Figure \ref{fig:mpm_sample} shows a sample simulation done using the MPM implementation. As is common in solutions to plasticity, an intermittent shear-band structure is seen to emerge surrounding the wheel, though the displacement itself appears smooth~\cite{ozaki2015finite}. 

\section{Results}

To begin, the performance of wheel C on PS prepared under various packing states is described. These experiments have two primary aims, first is to study the sensitivity of the RFT model to granular material density, and second is to analyze the capability of MPM-based continuum modeling in capturing system dynamics. Subsequently, the performance of wheels A and B on PS are presented. These experiments are aimed at investigating the ability of both the aforementioned methods in predicting the performance of wheels with diverse thickness--to--diameter aspect ratios. Finally, the performances of wheel C on MS and MMS sands are presented. These experiments are aimed at examining the capabilities of RFT to accurately model wheel performance when the force response surfaces for the granular material are not directly available, while also examining the capability of MPM continuum modeling for these cases.

Each experiment was performed at least five times, with the boxplots (Figures:~\ref{fig:WheelC_120_PS}, \ref{fig:Wheel_AB_PS}, \ref{fig:Wheel_C_MS_MMS} and \ref{fig:WheelC_130_MS}) showing the average and standard deviation. In order to quantify the performance of the various methods involved, several error metrics defined below were evaluated. Each of these metrics can help understand a particular aspect of the correlation between the model predictions and measured data. The metrics under consideration are the mean absolute error, the coefficient of correlation, and the coefficient of variation. The mean absolute error $\Delta$ is defined as follows:

\begin{equation}
\Delta = \frac{1}{k}\sum_i^k |X_e - X_m|_i
\end{equation}
where $X_e$ is the experimental average (either traction $F_x$, torque $M$, or sinkage $z$), $X_m$ is the model prediction, and $k$ is the number of data points used in the evaluation. The mean absolute error provides an estimate of the absolute deviations, and has the dimensions of the quantity under investigation. 

The coefficient of correlation $R$ is used to evaluate the correlation between the trends of the modeled predictions and the measured values. The coefficient of correlation $R$ is defined as
\begin{equation}
R = \frac{k \sum_i^k X_e X_m - \sum_i^k X_e \sum_i^n X_m}{\left[ \sqrt{k \sum_i^k X_e^2-\left(\sum_i^k X_m \right)^2 } \right]  \left[ \sqrt{k \sum_i^k X_m^2-\left(\sum_i^k X_e \right)^2} \right] }
\end{equation}

A value of 1.0 for the coefficient of correlation $R$, indicates a perfect correlation between the trends of the predicted and measured data. The correlation is generally regarded as strong if the value of $R$ is greater than 0.8. With a value of $R$ less than 0.5, the correlation is usually regarded as weak. Finally, the coefficient of variation $CV$ is defined as follows:
\begin{equation}
CV = \sqrt{\frac{\sum_i^k \left(X_e-X_m\right)^2}{k \sum_i^k (X_e)^2}}\,,
\end{equation}
Where, $X_e$ and $X_m$ are experimental and model predicted values respectively and k represents the total number of slip values at which experiments are done for a given load value. The $CV$ provides a normalized measure of deviations. If the value of $CV$ is zero, the predicted and measured data will have a perfect match, representing a zero deviation between model and experiment.

\subsection{Sensitivity to Poppy Seeds Packing State}

\begin{figure}[htbp]
\centering
\subfigure[]{
\includegraphics[trim = 1mm 10mm 10mm 5mm, clip, width = 0.47\textwidth]{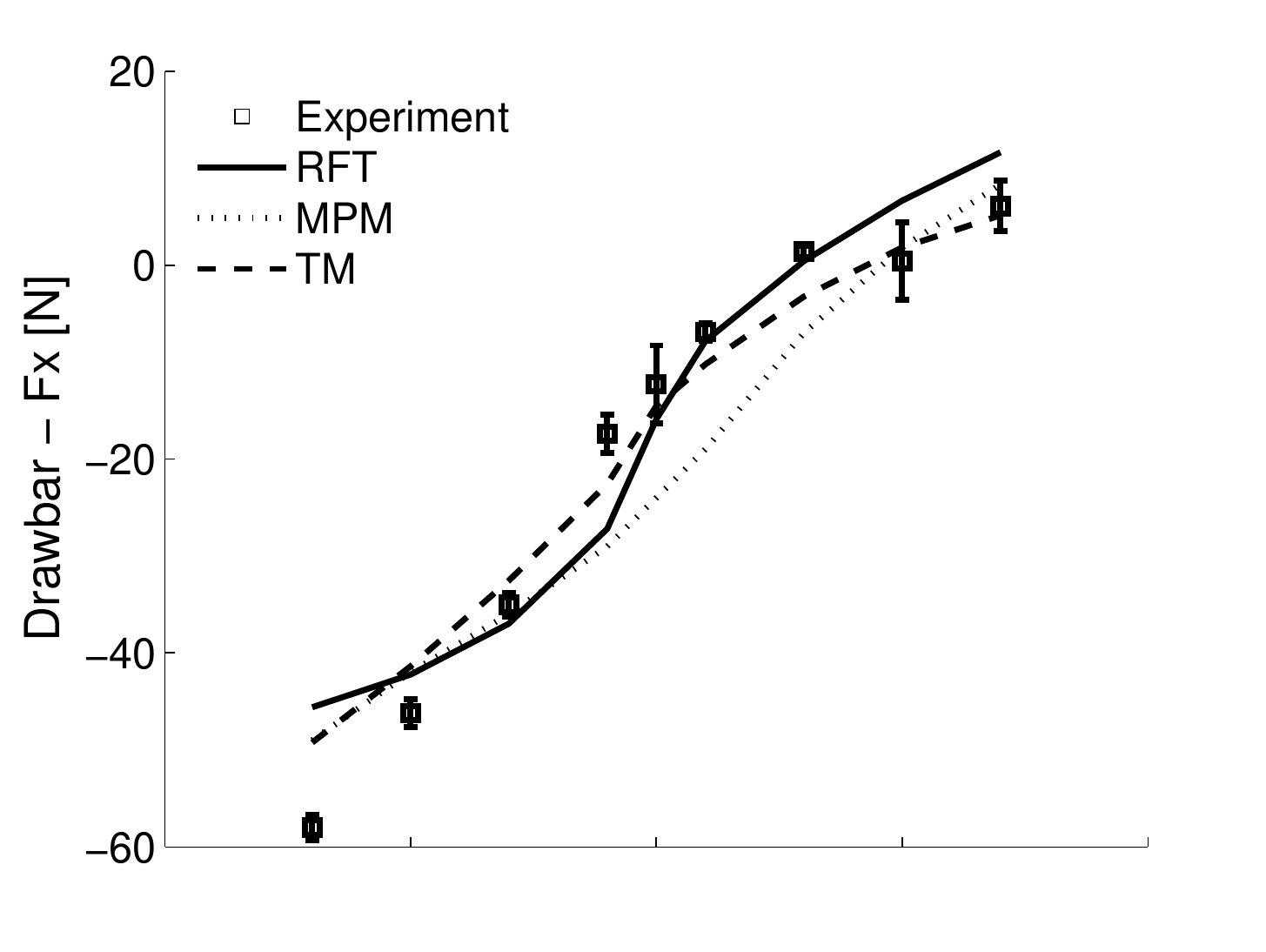}
\label{fig:DP_WheelC_120_CPS}
}
\subfigure[]{
\includegraphics[trim = 1mm 10mm 10mm 5mm, clip, width = 0.47\textwidth]{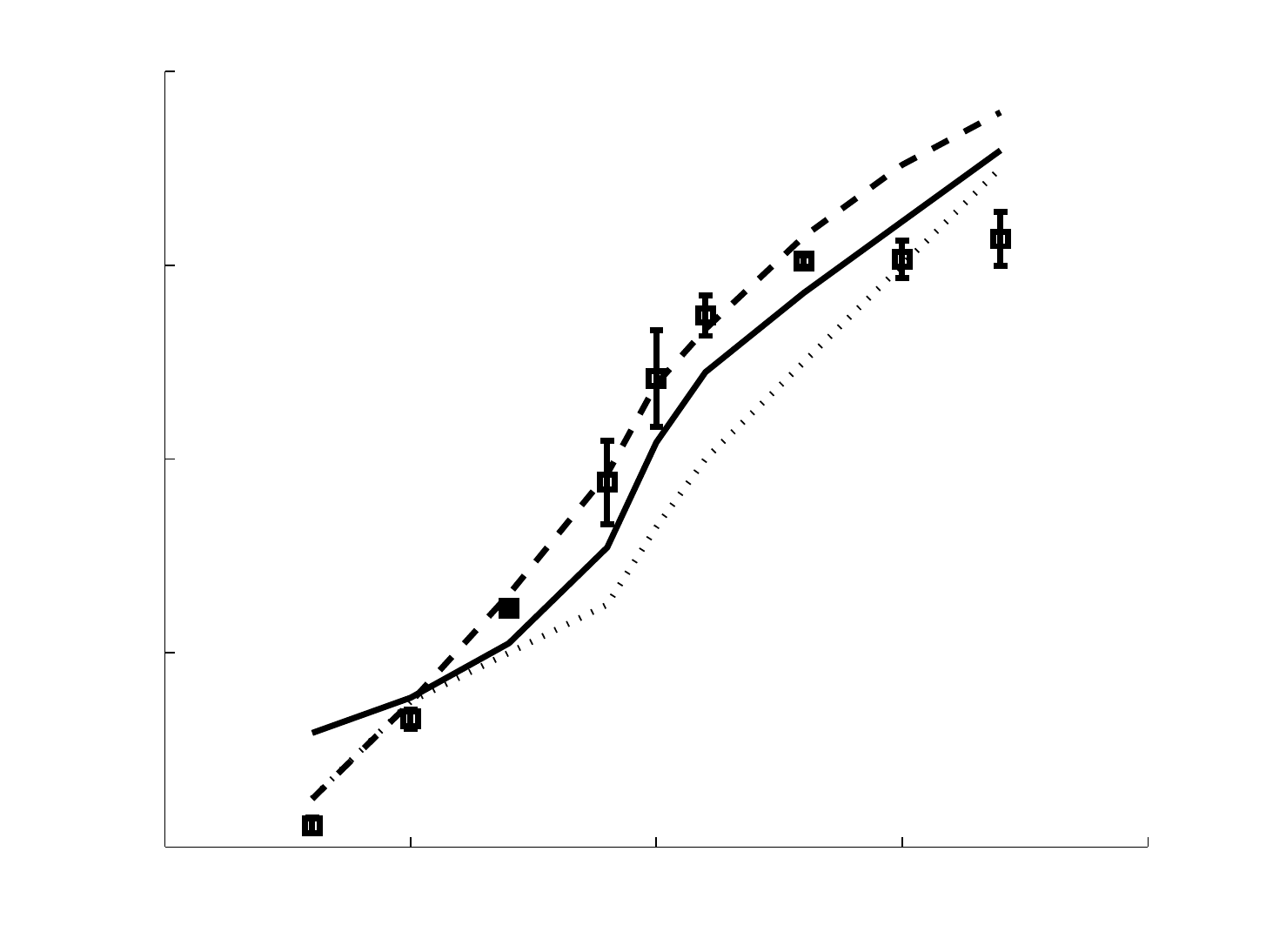}
\label{fig:DP_WheelC_120_LPS}
}
\subfigure[]{
\includegraphics[trim = 1mm 10mm 10mm 5mm, clip, width = 0.47\textwidth]{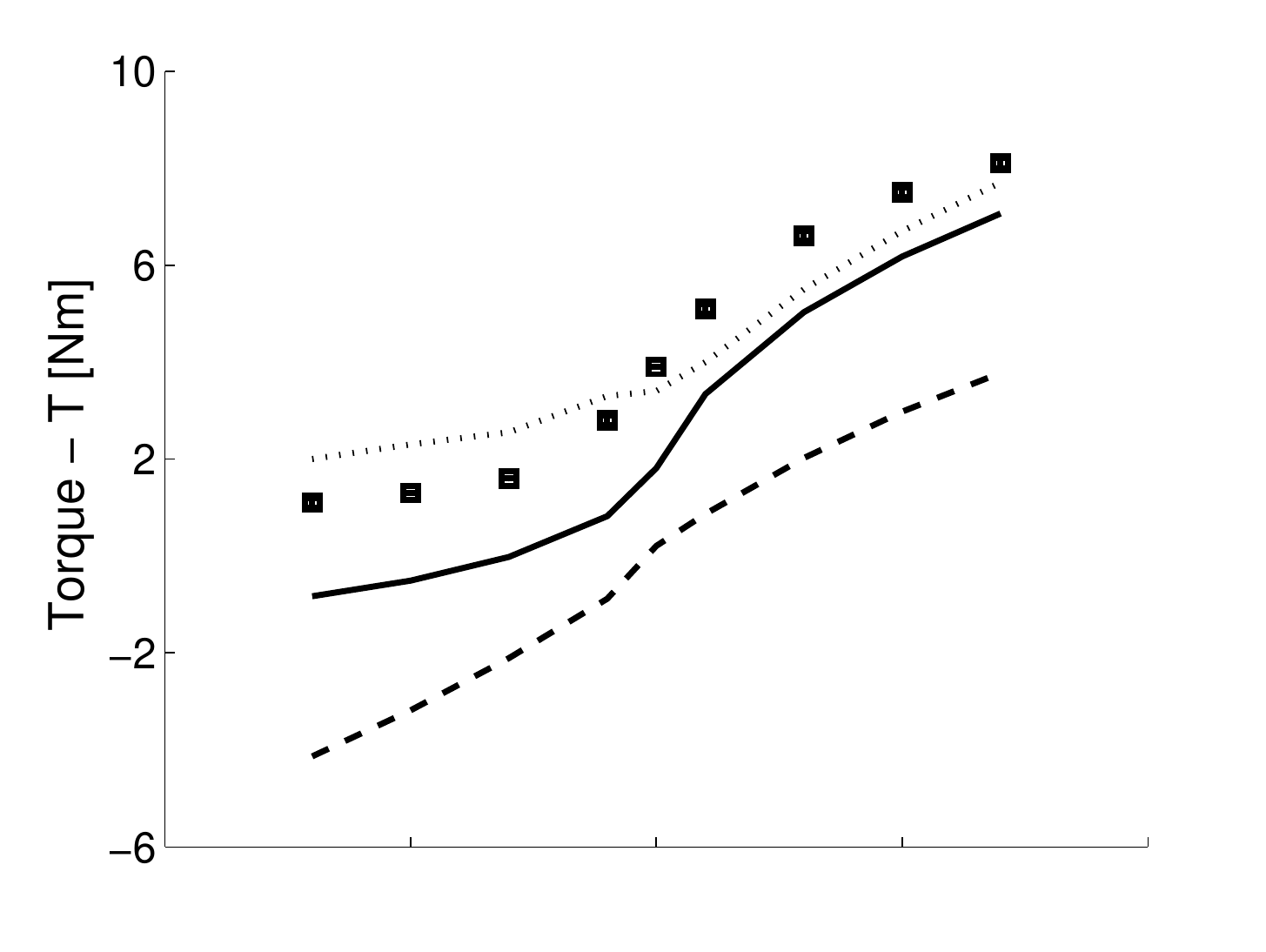}
\label{fig:T_WheelC_120_CPS}
}
\subfigure[]{
\includegraphics[trim = 1mm 10mm 10mm 5mm, clip, width = 0.47\textwidth]{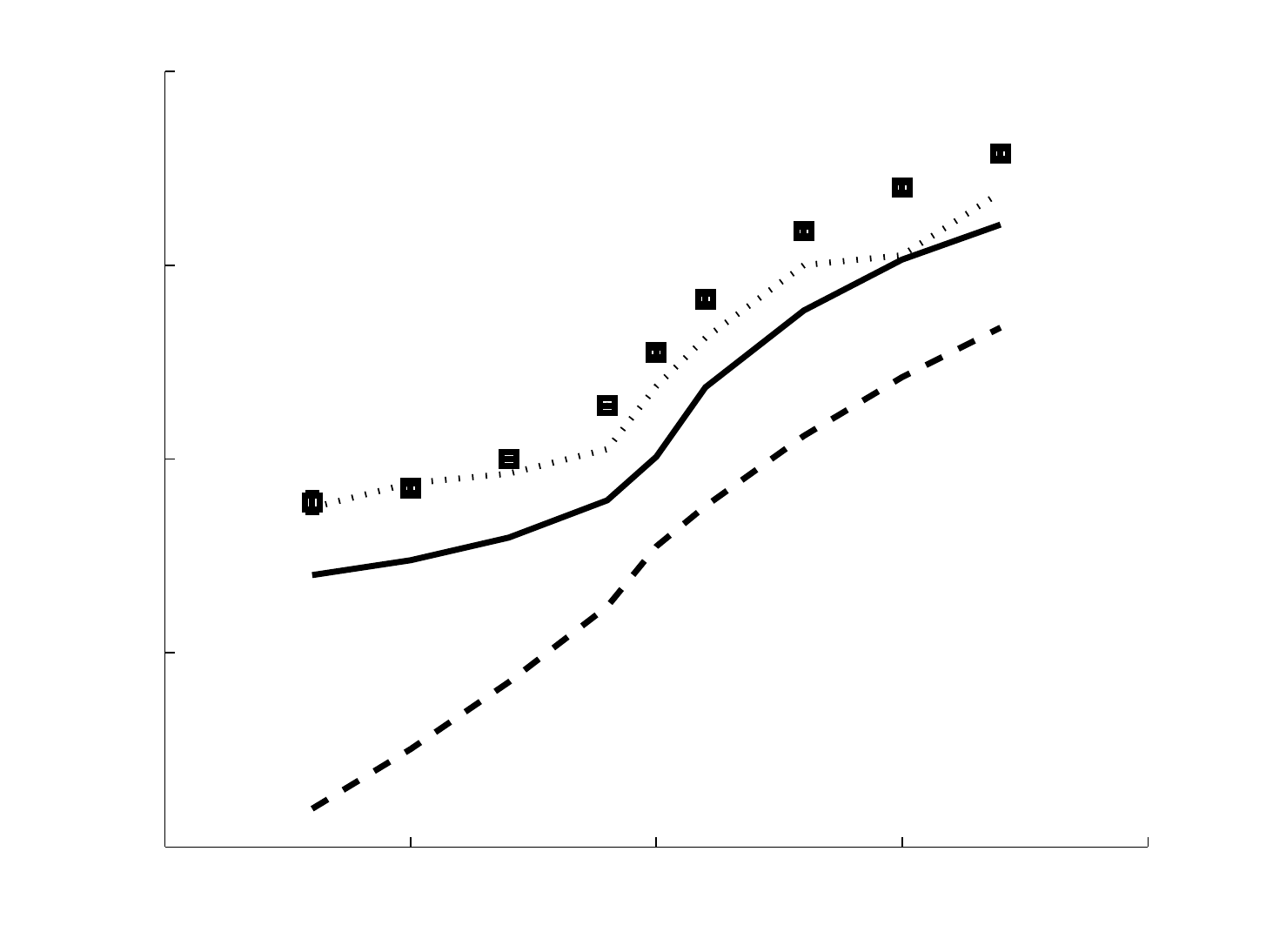}
\label{fig:T_WheelC_120_LPS}
}
\subfigure[]{
\includegraphics[trim = 1mm 1mm 10mm 5mm, clip, width = 0.47\textwidth]{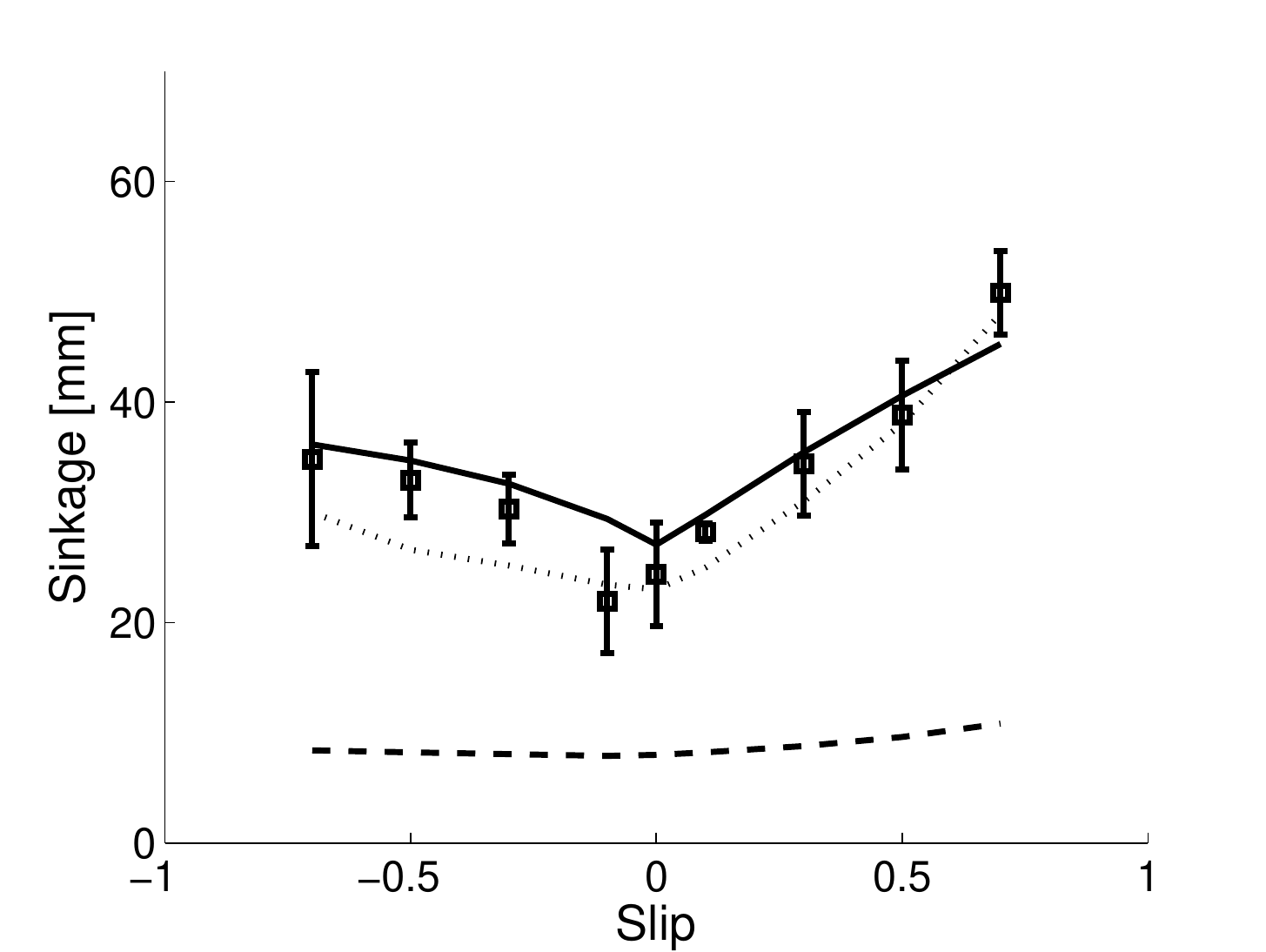}
\label{fig:Z_WheelC_120_CPS}
}
\subfigure[]{
\includegraphics[trim = 1mm 1mm 10mm 5mm, clip, width = 0.47\textwidth]{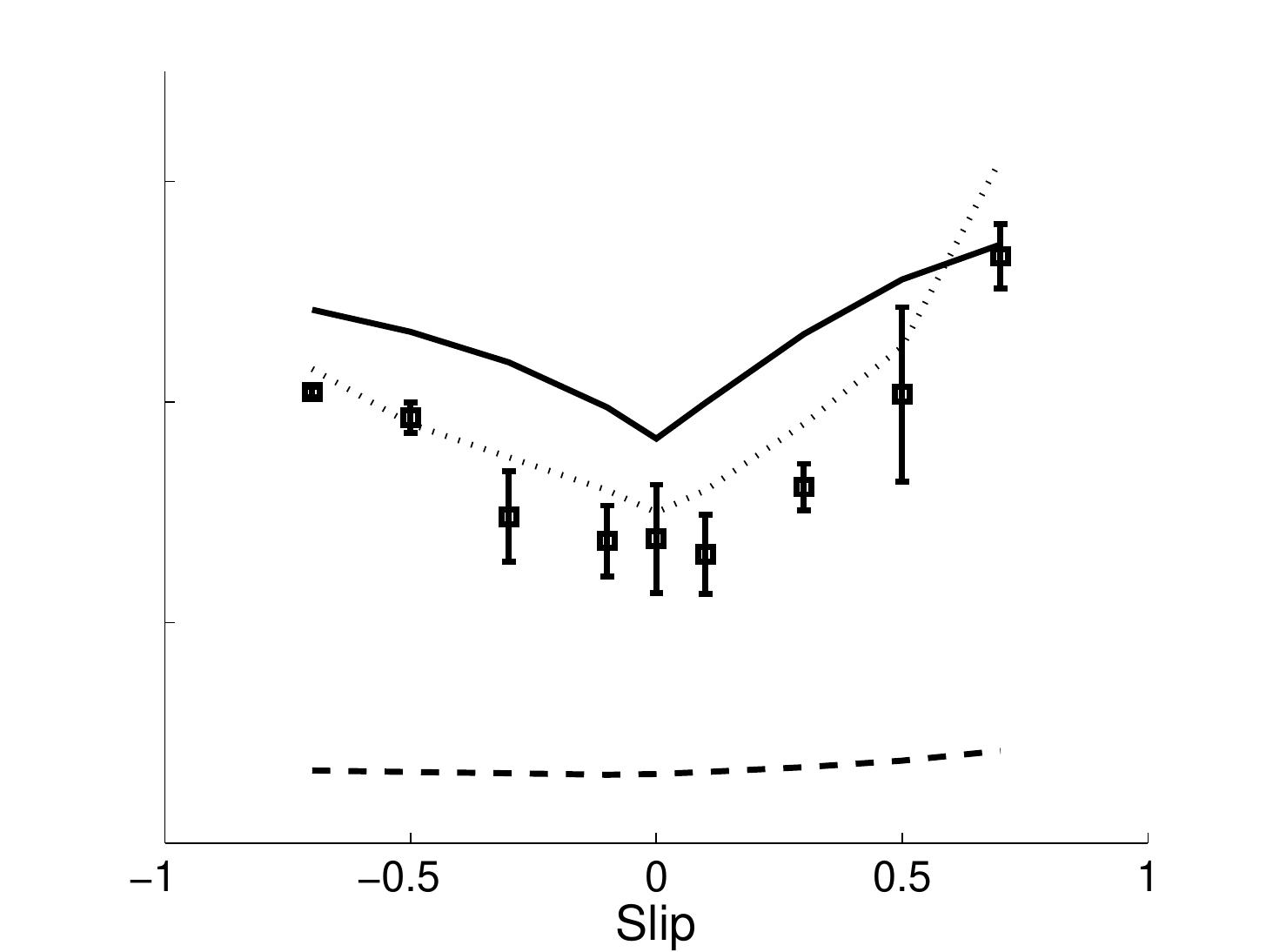}
\label{fig:Z_WheelC_120_LPS}
}
\caption{Wheel C on Poppy Seeds. (a), (c) and (e) correspond to the dense poppy seed state ($\phi$ $\approx$ 0.60) while (b), (d) and (f) correspond to the loose poppy seed state ($\phi$ $\approx$ 0.58). Experiments were performed five times and boxplots present the average reading and one standard deviation. Nominal vertical load is 120 N. Resistive force theory (RFT), Continuum modeling(MPM) and terramechanics (TM) approaches produce similar predictions for drawbar, while the RFT and MPM outperform the TM model when sinkage (at the dense state) is evaluated. Torque predictions show visible deviation for all the models with the RFT and MPM producing estimates closer to measured values.}
\label{fig:WheelC_120_PS}
\end{figure}

Figure \ref{fig:WheelC_120_PS} presents experimental results of the wheel C on PS. To highlight potential quantitative differences in wheel performance during travel on soil in loose and compact states, the granular material was prepared at two packing fractions that were chosen to span the onset of dilatancy. For the packing states selected, the poppy seeds show different behavior under plate penetration tests, which results in different RFT properties. However, experiments show that wheel performance is moderately affected by terrain preparation. Drawbar measurements for the loose and dense states are within 25\% of each other. In absolute terms, the difference between loose and dense packing does not exceed 4 N for any tested slip level. Torque measurements stay within a 7\% difference, while the sinkages' average variation is 11\%. 

The fact that wheel performance is unaffected by terrain preparation is surprising, since on firmer terrain one would expect less sinkage and thus increased traction. The high difference in angle of repose of LPS and CPS confirms large differences in initial medium state; the small difference in wheel performance is surprising. It is possible that this is a result peculiar to the poppy seeds' mechanical properties or it could be the low penetration of wheel into medium which causes these effects. 

\begin{table}[htbp]
\centering
\caption{Comparison of resistive force theory (RFT), continuum modeling (MPM) and terramechanics (TM) models' predictions for wheel C on Poppy Seeds under 120 N nominal load. The mean absolute error $\Delta$ has dimension of [N] for drawbar, [Nm] for torque, and [mm] for sinkage. The coefficient of correlation $R$ and the coefficient of variation $CV$ are unitless.}
\centering
\vspace{0.5em}
\begin{tabular}{c|ccc|ccc}
&	\multicolumn{3}{c}{Compacted}	&	\multicolumn{3}{{c}}{Loose} \\ 
\hline
& RFT & MPM & TM & RFT & MPM & TM  \\ \hline
&	\multicolumn{6}{c}{Drawbar} \\ 
$\Delta$	&	5.05	&	6.84	& 3.41	&	5.64	&	7.80	& 3.58	\\	
$R$			&	0.92	&	0.94 	& 0.99  &	0.96	&	0.93	& 0.99		\\
$CV$		&	0.08	&	0.12 	& 0.08  &	0.07	&	0.11	& 0.05		\\
\hline

&	\multicolumn{6}{c}{Torque} \\ 
$\Delta$	&	1.68	&	0.81	& 4.30	&	1.68	&	0.64	& 4.46		\\	
$R$			&	1.00	&	0.97 	& 0.98	&	1.00	&	0.99	& 0.99		\\
$CV$		&	0.32	&	0.16 	& 0.78 	&	0.33	&	0.15	& 0.88		\\
\hline
&	\multicolumn{6}{c}{Sinkage} \\
$\Delta$	&	2.73	&	3.16	& 26.07	&	9.94	&	4.41	& 26.48		\\	
$R$			&	0.97	&	0.96 	& 0.93	&	0.93	&	0.96	& 0.86		\\
$CV$		&	0.11	&	0.10 	& 0.80	&	0.30	&	0.14	& 0.76		\\

\end{tabular}
\label{tab:PS_C_statistics}
\end{table}

Table \ref{tab:PS_C_statistics} presents the values of mean absolute error, coefficient of correlation, and coefficient of variation for RFT, MPM, and TM models. Excepting drawbar outcomes, RFT and MPM consistently show lower mean absolute error values, high coefficient of correlation values, and lower coefficient of variation values than TM. In particular, the RFT performance improves when high density terrain parameters are used, especially when sinkage is considered. The high coefficients of correlation shows that all the models follow the trends of experimental data. The lower coefficient of variation highlights how MPM performs better than the other two models especially for torque measurement. 

It should be reiterated that the MPM simulations were conducted assuming plane strain conditions in the wheel locomotion. This approximation is less valid at high wheel sinkages due to the reduced aspect ratio of the wheel--media interface area, hence an exact match of results for high sinkage cases is not expected. The terrain parameters for the TM model (only for PS) were also not calculated according to standard terramechanics practices. According to terramechanics guidelines, the dimensions of the intruder used for finding TM fitting parameters should approximately be the same as the average contact patch area of the wheels. But in the above analysis, the wheels used had a much different contact patch area than that of intruder ($2.5 \times 3.8 $ $\rm{ cm^2}$). Hence, this could partially explain the poor performance shown by the TM model. In order to obtain meaningful drawbar predictions, the shear displacement modulus was set to 0.04 m which is larger (by a factor of two) than any value found in the literature. A large shear modulus means that larger deformations are needed to generate shear stress which can be consistent with the nature of poppy seeds.

\subsection{Sensitivity to Wheel Geometry on Poppy Seeds}

\begin{figure}[htbp]
\centering
\subfigure[]{
\includegraphics[trim = 1mm 10mm 10mm 5mm, clip, width = 0.47\textwidth]{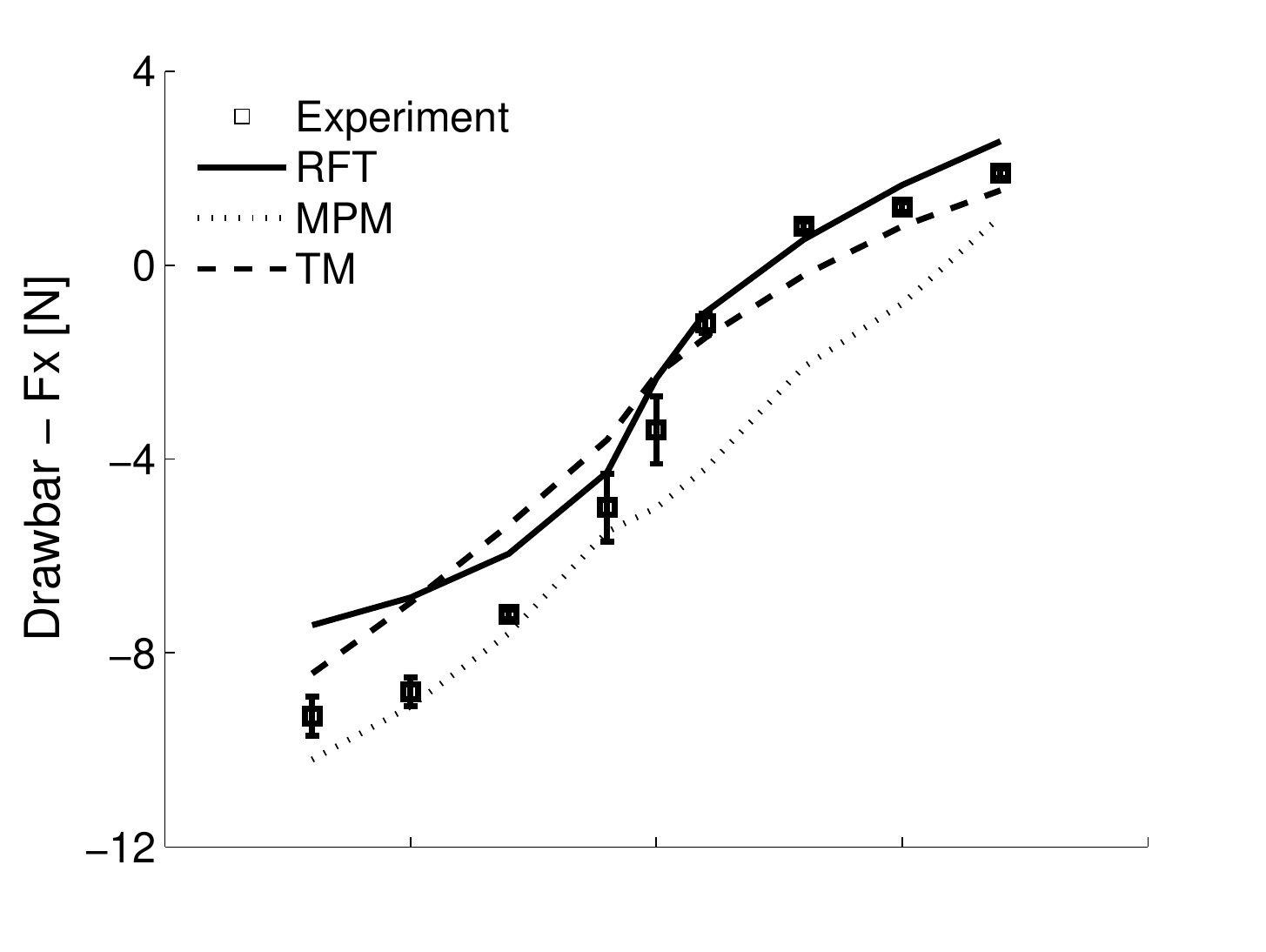}
\label{fig:WheelA_20_DP}
}
\subfigure[]{
\includegraphics[trim = 1mm 10mm 10mm 5mm, clip, width = 0.45\textwidth]{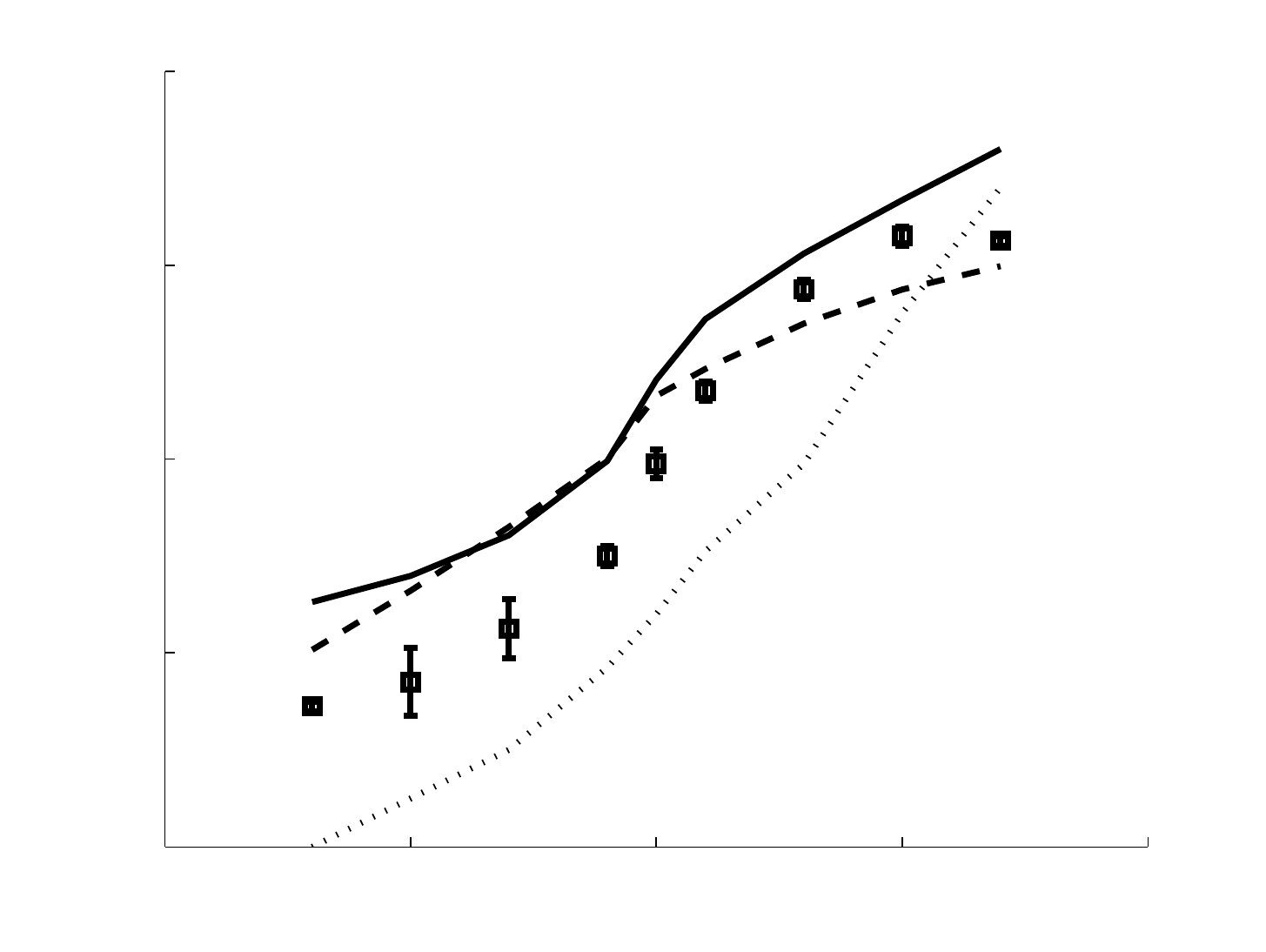}
\label{fig:WheelB_18_DP}
}
\subfigure[]{
\includegraphics[trim = 1mm 10mm 10mm 5mm, clip, width = 0.45\textwidth]{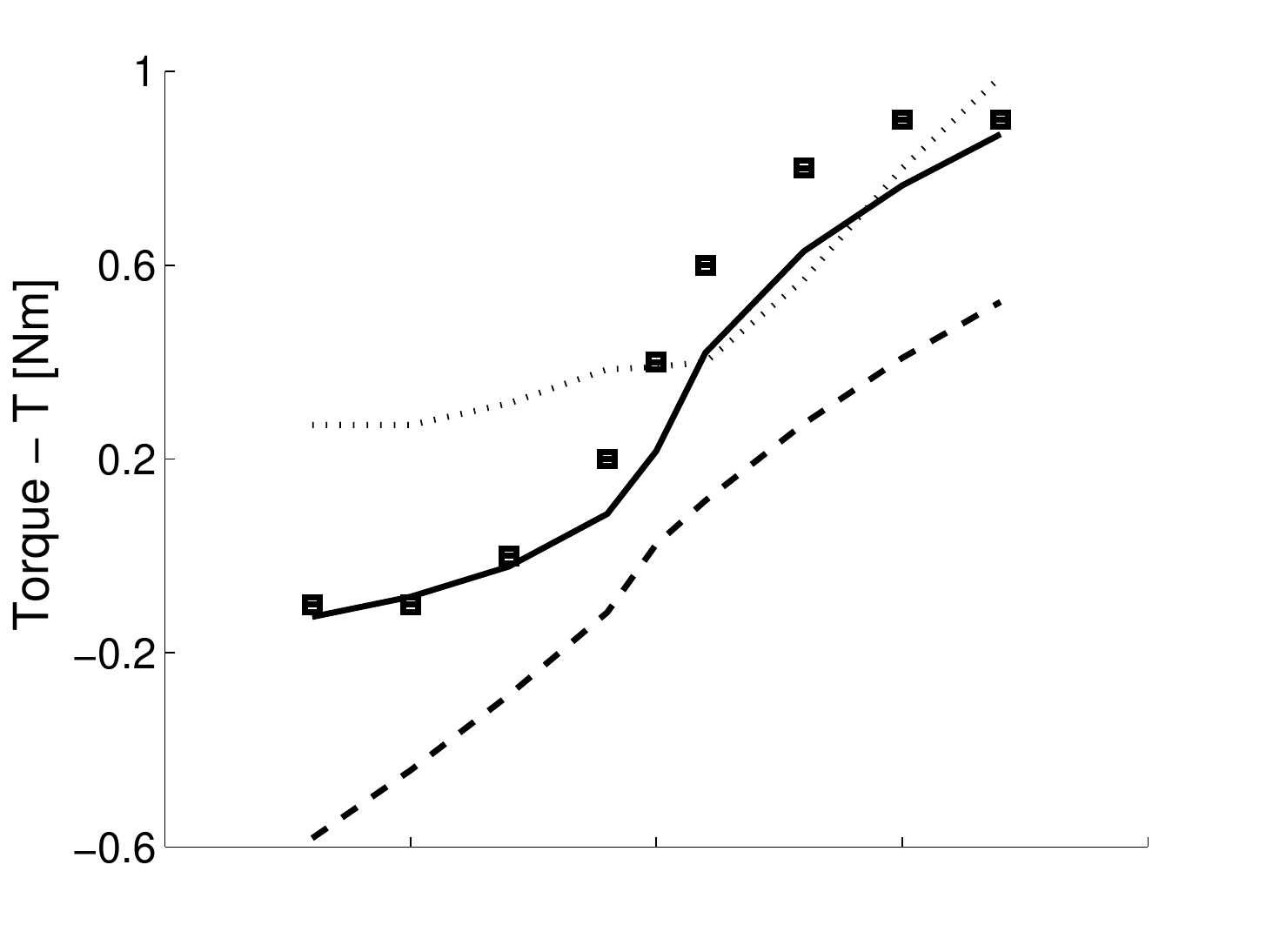}
\label{fig:WheelA_20_T}
}
\subfigure[]{
\includegraphics[trim = 1mm 1mm 10mm 5mm, clip, width = 0.45\textwidth]{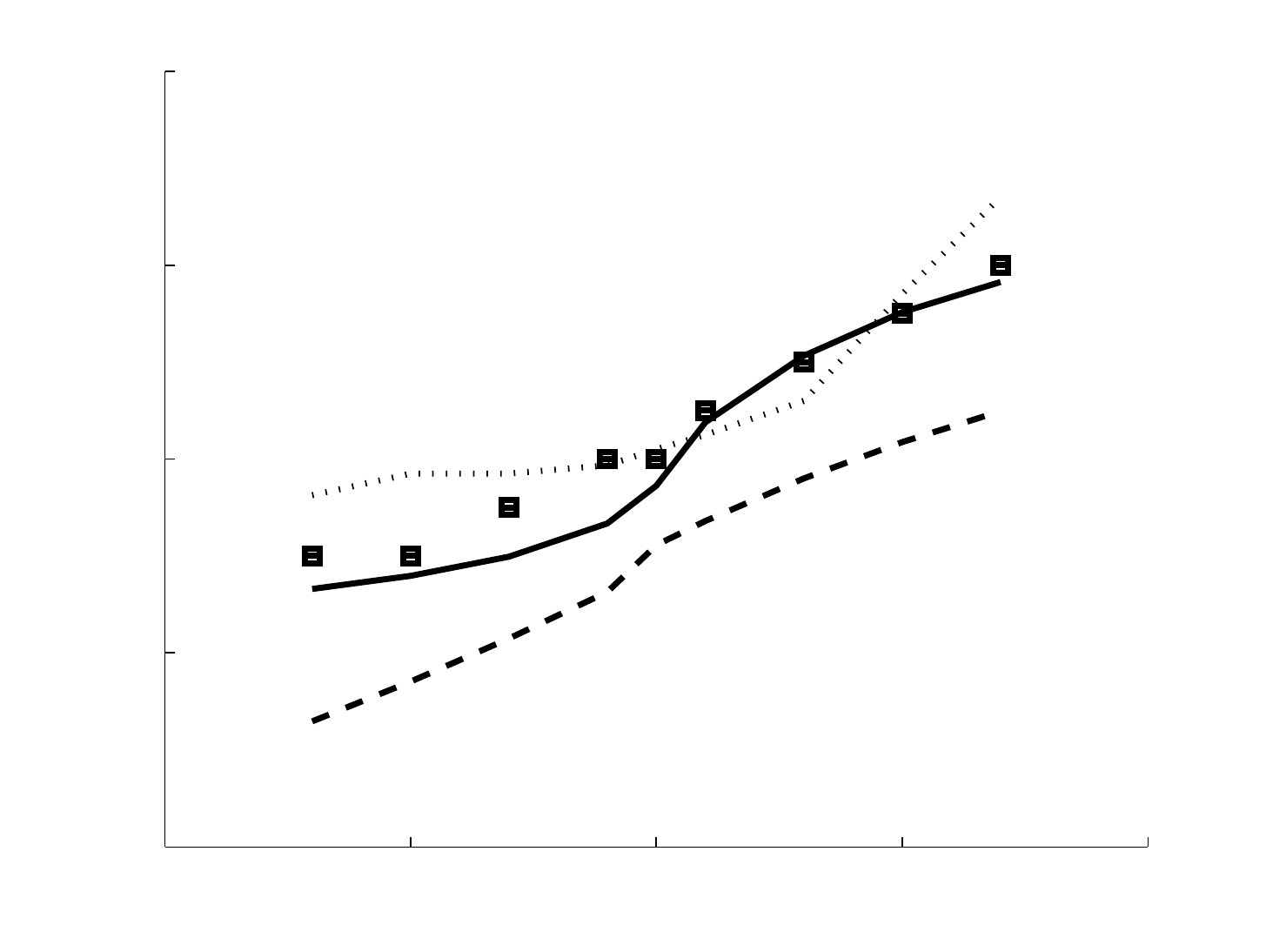}
\label{fig:WheelB_18_T}
}
\subfigure[]{
\includegraphics[trim = 1mm 1mm 10mm 5mm, clip, width = 0.45\textwidth]{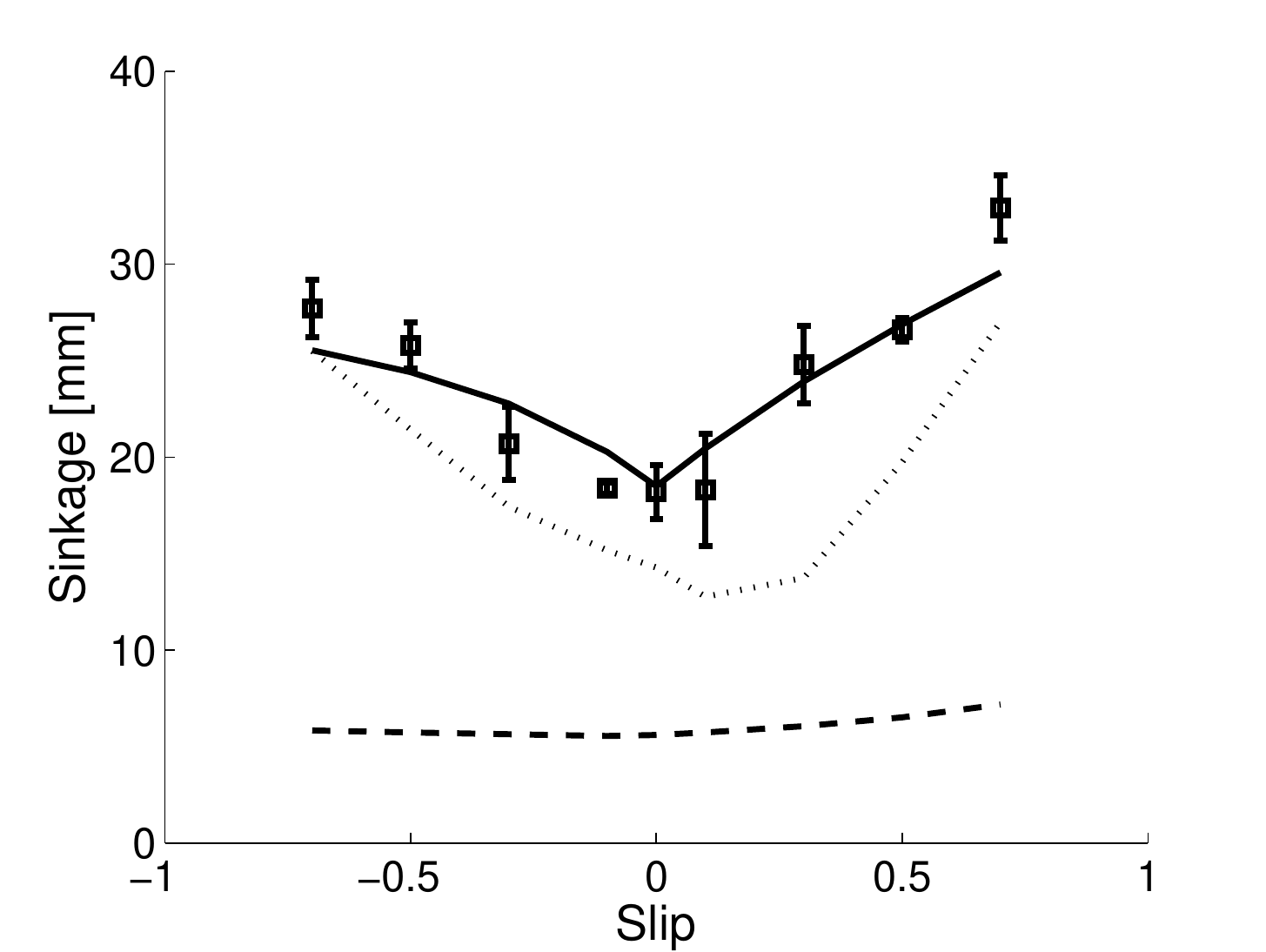}
\label{fig:WheelA_20_Z}
}
\subfigure[]{
\includegraphics[trim = 1mm 1mm 10mm 5mm, clip, width = 0.47\textwidth]{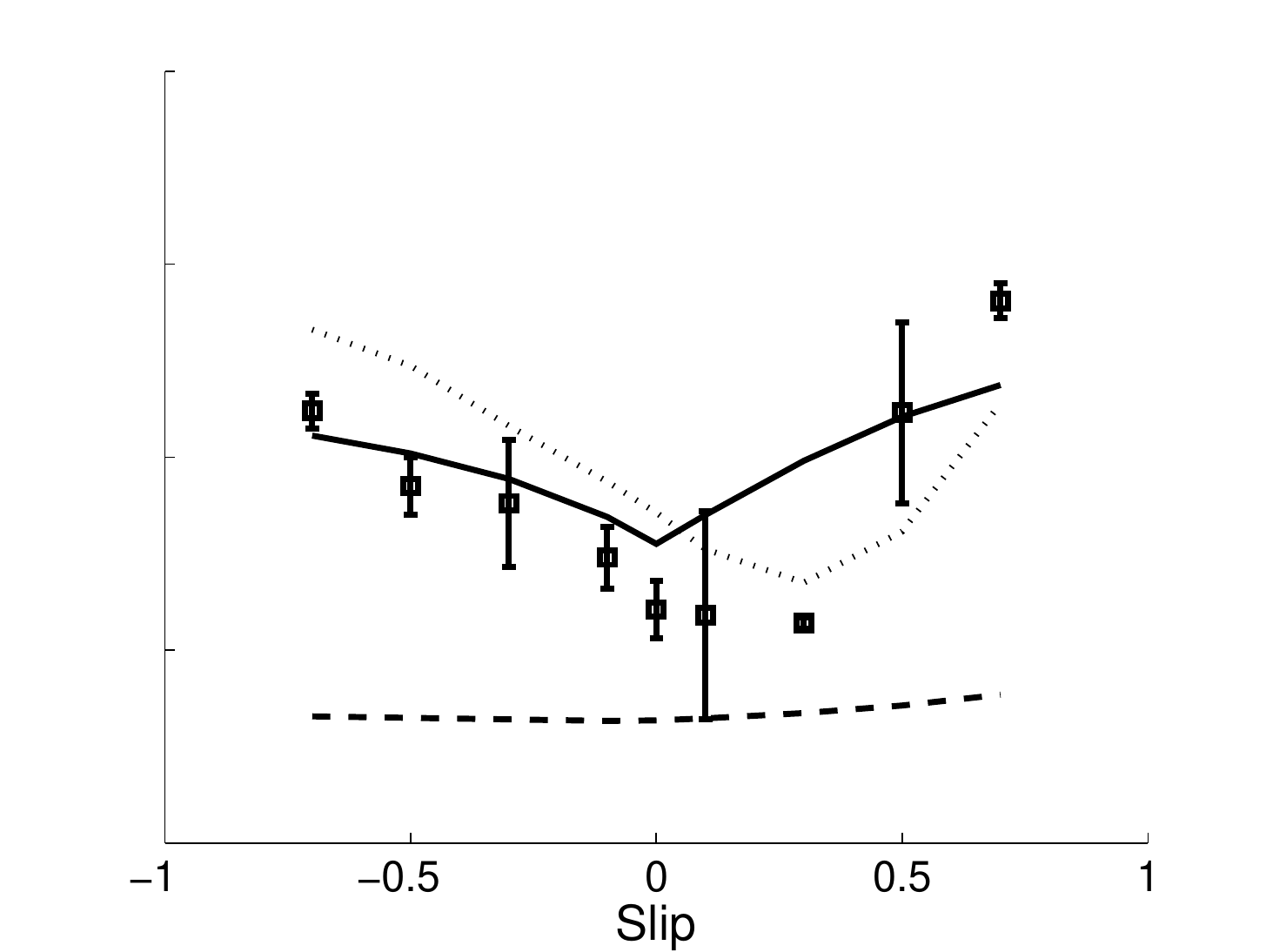}
\label{fig:WheelB_18_Z}
}
\caption{Wheel A (a,c,e) and wheel B (b,d,f) on dense poppy seeds. Experiments were performed five times and boxplots present the average reading and one standard deviation. Nominal vertical load is 20 N for wheel A, and 18 N for wheel B. All experiments are conducted for a packing state of $\phi$ = 0.60. Resistive force theory (RFT), continuum modeling (MPM) and terramechanics (TM) approaches produce similar predictions for drawbar, while the RFT outperforms MPM and the TM model when sinkage is evaluated. Torque predictions show visible deviation for all the models with RFT and MPM producing estimates closer to measured values.
}
\label{fig:Wheel_AB_PS}
\end{figure}

Figure \ref{fig:Wheel_AB_PS} presents the results obtained with wheels A and B on dense poppy seeds. These wheels have different aspect ratios and geometries, with wheel B being a lugged wheel and wheel A being a smooth wheel. 
Table \ref{tab:PS_AB_statistics} presents the values of mean absolute error, coefficient of correlation, and coefficient of variation for the RFT, continuum model (MPM), and the TM model. For all the outputs considered here, RFT consistently shows lower mean absolute error, higher coefficient of correlation, and lower coefficient of variation values than the TM model for torque and sinkage.    

\begin{table}[htbp]
\centering
\caption{Comparison of resistive force theory (RFT), continuum modeling(MPM) and terramechanics (TM) predictions for wheels A and B on Poppy Seeds under 20 N and 18 N nominal load respectively. The mean absolute error $\Delta$ has dimension of [N] for drawbar, [Nm] for torque, and [mm] for sinkage. Coefficients of correlation $R$ and coefficients of variation $CV$ are unitless.}
\centering
\vspace{0.5em}

\begin{tabular}{c|ccc|ccc}
&	\multicolumn{3}{c}{A}	&	\multicolumn{3}{{c}}{B} \\ 
\hline
& RFT & MPM & TM & RFT & MPM & TM  \\ \hline
&	\multicolumn{6}{c}{Drawbar} \\ 
$\Delta$	&	0.94	&	1.39	& 1.02	&	1.65	&	2.53	&	1.26	\\	
$R$			&	1.00	&	0.98	& 0.99	&	0.99	&	0.96 	&	0.98	\\
$CV$		&	0.07	&	0.11	& 0.07	&	0.10	&	0.16 	&	0.08	\\
\hline

&	\multicolumn{6}{c}{Torque} \\ 
$\Delta$	&	0.10	&	0.21	& 0.41	&	0.05	&	0.08	&	0.26	\\	
$R$			&	0.99	&	0.87	& 0.98	&	0.99	&	0.91 	&	0.98	\\
$CV$		&	0.21	&	0.43	& 0.74	&	0.20	&	0.29 	&	0.82	\\
\hline
&	\multicolumn{6}{c}{Sinkage} \\
$\Delta$	&	1.60	&	5.15	& 17.73	&	3.10	&	4.49	&	10.98	\\	
$R$			&	1.00	&	0.87	& 0.83	&	0.87	&	0.64 	&	0.75	\\
$CV$		&	0.08	&	0.24	& 0.76	&	0.21	&	0.25 	&	0.66	\\

\end{tabular}
\label{tab:PS_AB_statistics}
\end{table}

On the other hand, considering drawbar force, TM performs better than RFT (though the difference is not high with both methods having CV below 0.10 and R above 90\%). Thus based on requirement of high R and low CV, it can be concluded that, for the cases considered here, in general RFT shows better performance than the TM model. 

The performance of MPM appears to be on par with the TM model in all of the above cases. Considering torque, while the CV values are high, the absolute value of mean error is within 0.3 Nm. Sinkage comparisons show a better performance (lower mean absolute error) of MPM than TM, but the correlation coefficient values were low. As mentioned before, performing continuum simulations for wheels of narrow aspect ratios under plane-strain conditions is another possible source of error.

\subsection{Sensitivity to Vertical Load on MS and MMS sands}

\begin{figure}[htbp]
\centering
\subfigure[]{
\includegraphics[trim = 1mm 10mm 10mm 5mm, clip, width = 0.47\textwidth]{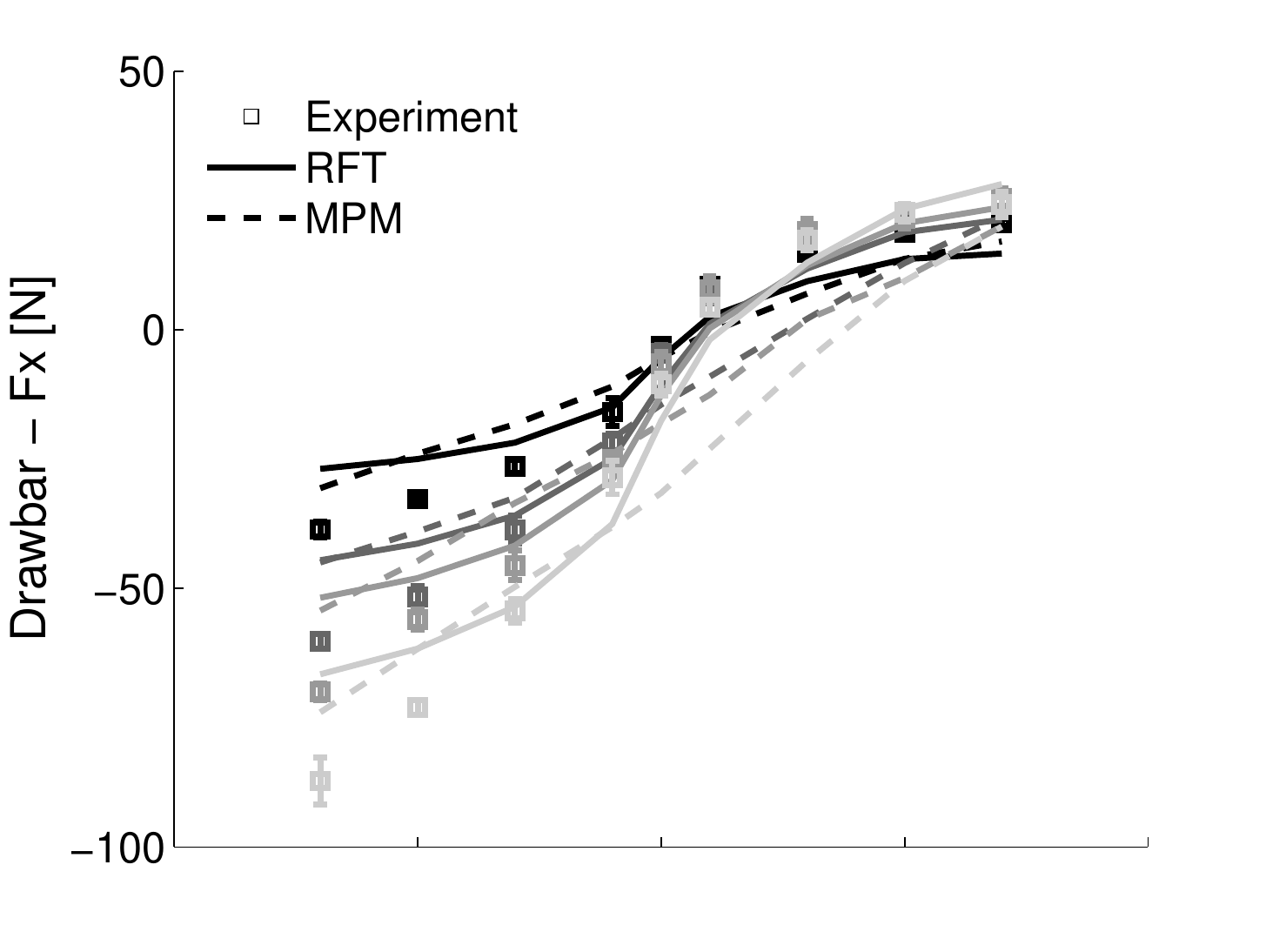}
\label{fig:WheelC_MS_DP}
}
\subfigure[]{
\includegraphics[trim = 1mm 10mm 10mm 5mm, clip, width = 0.47\textwidth]{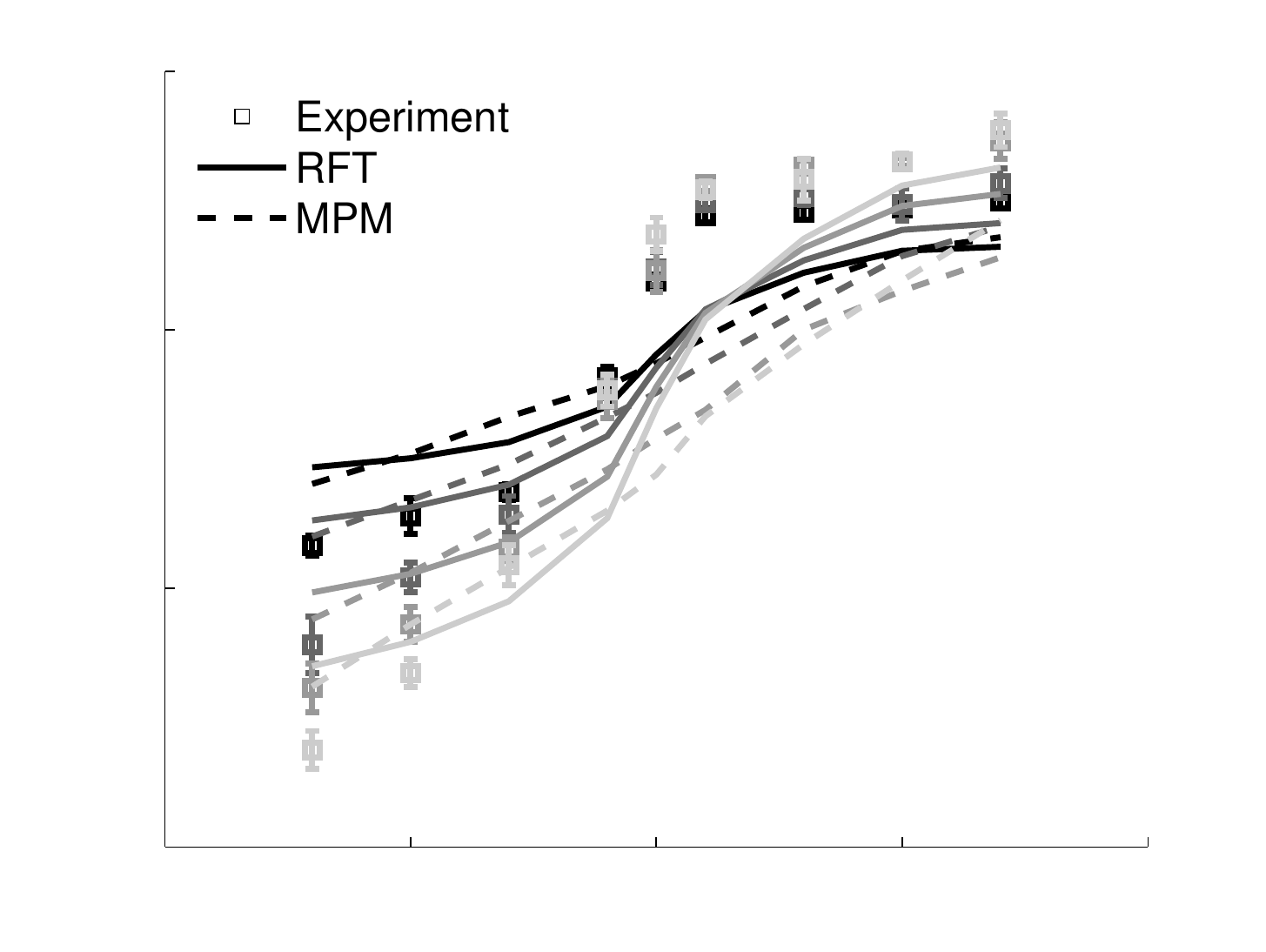}
\label{fig:WheelC_MMS_DP}
}

\subfigure[]{
\includegraphics[trim = 1mm 10mm 10mm 5mm, clip, width = 0.47\textwidth]{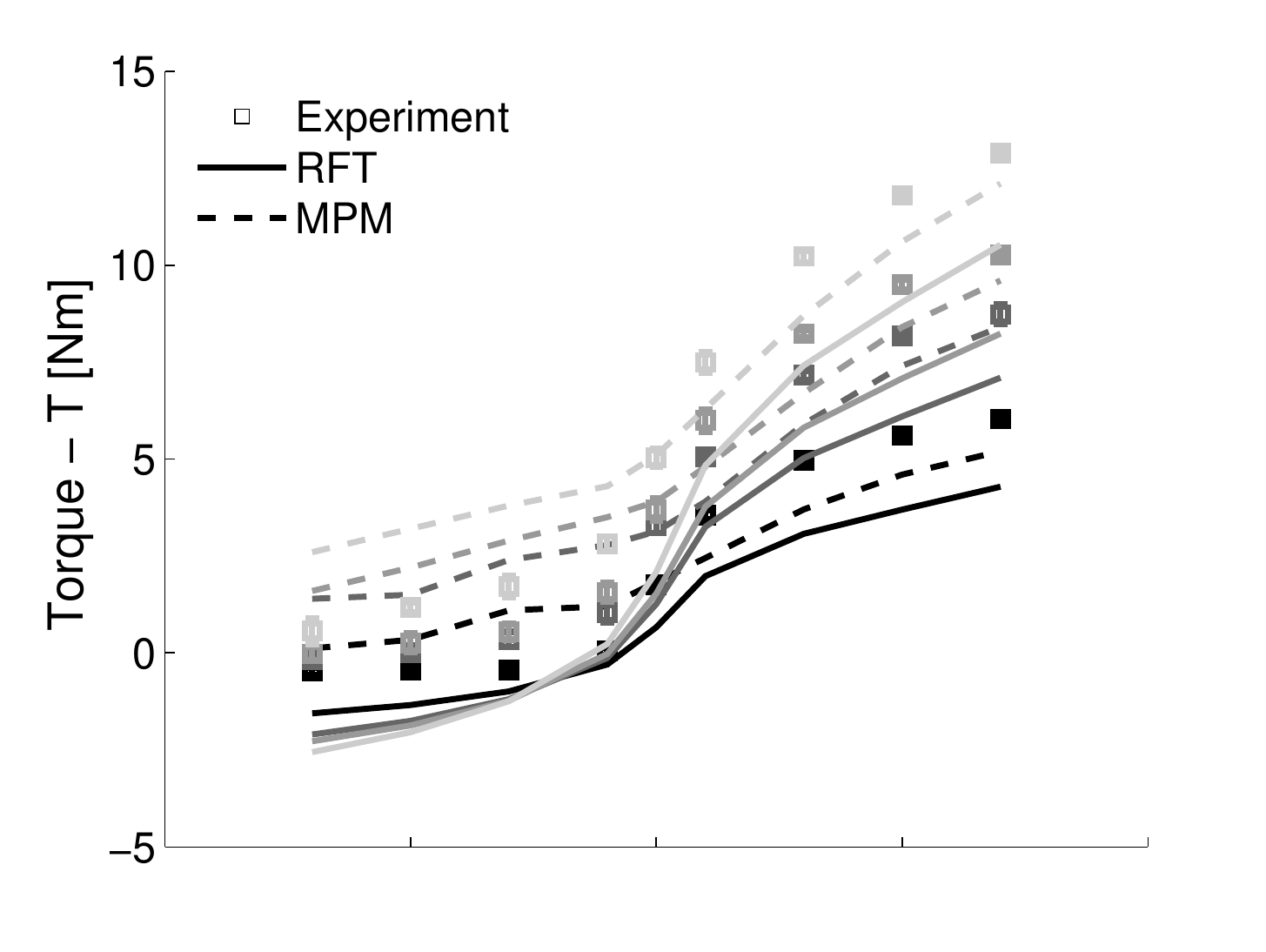}
\label{fig:WheelC_MS_T}
}
\subfigure[]{
\includegraphics[trim = 1mm 10mm 10mm 5mm, clip, width = 0.47\textwidth]{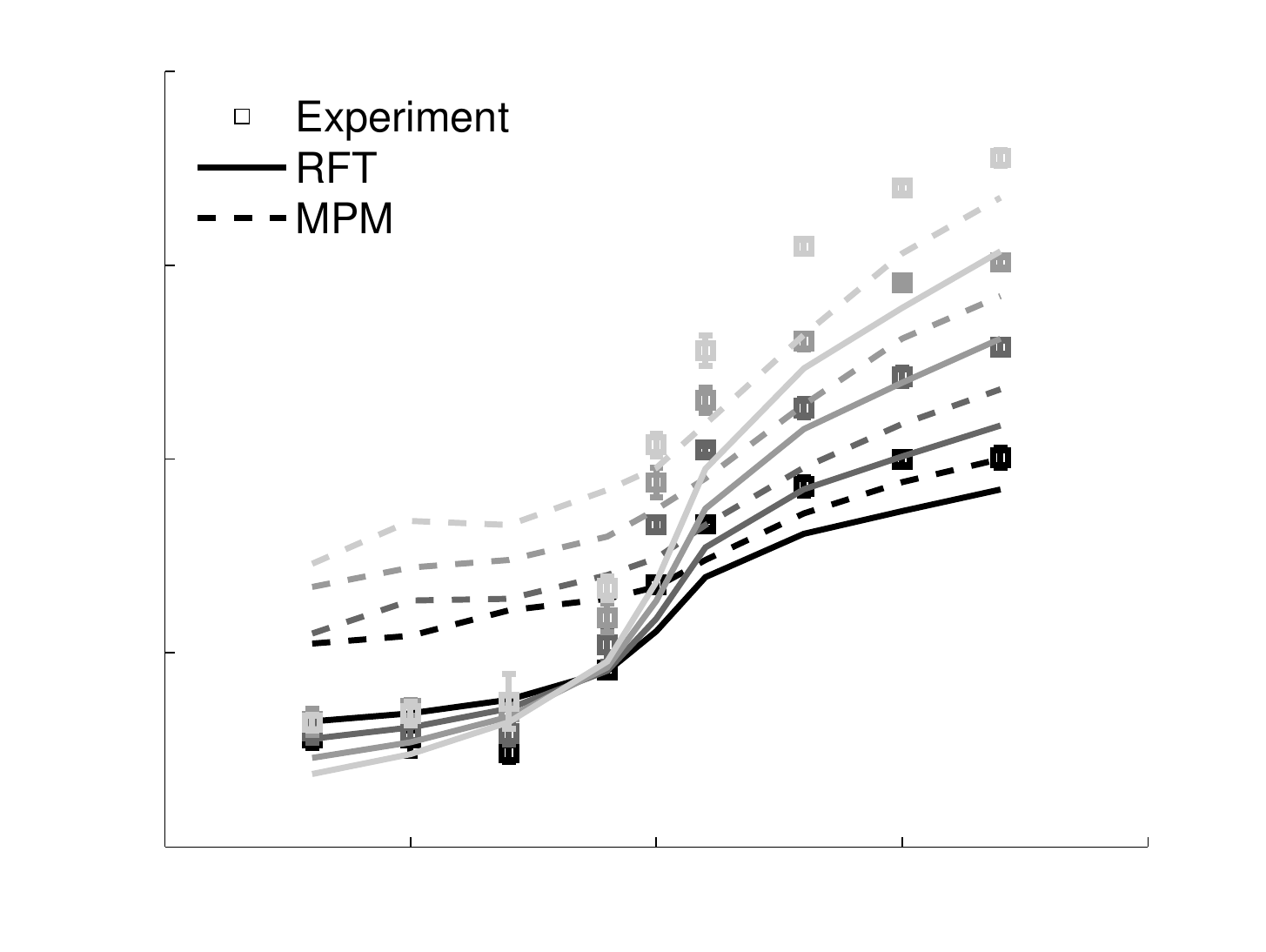}
\label{fig:WheelC_MMS_T}
}

\subfigure[]{
\includegraphics[trim = 1mm 1mm 10mm 5mm, clip, width = 0.47\textwidth]{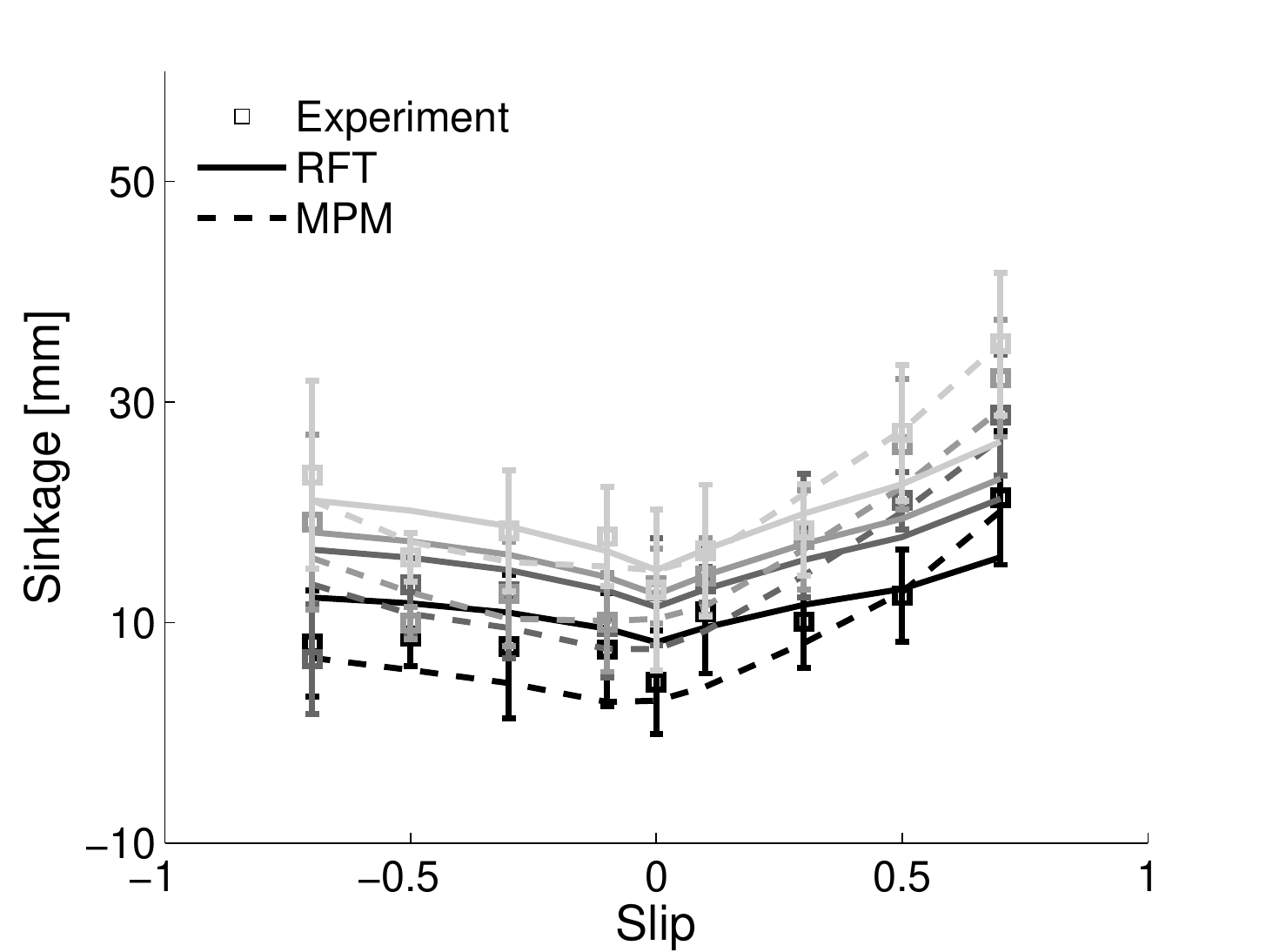}
\label{fig:WheelC_MS_Z}
}
\subfigure[]{
\includegraphics[trim = 1mm 1mm 10mm 5mm, clip, width = 0.47\textwidth]{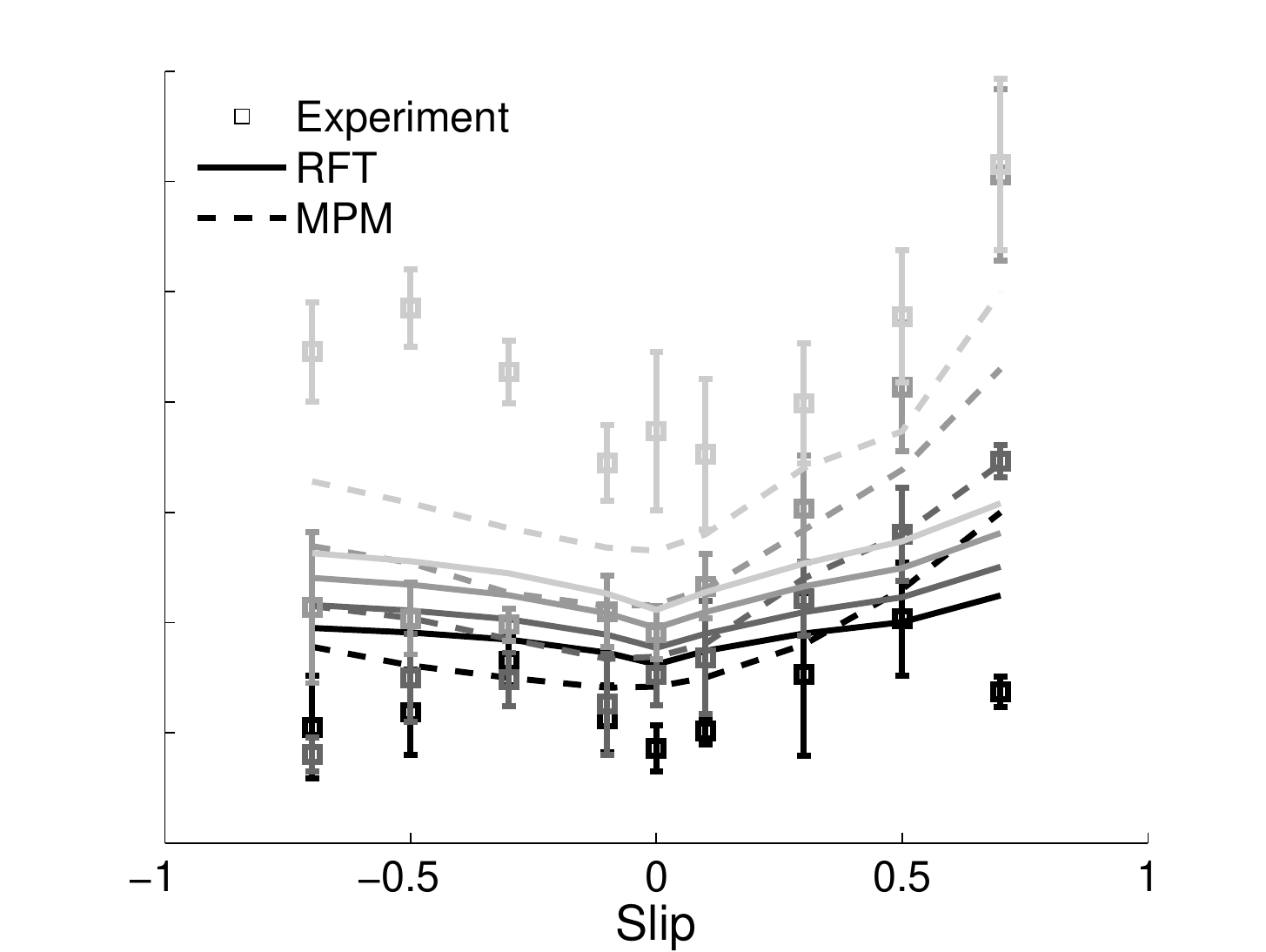}
\label{fig:WheelC_MMS_Z}
}

\caption{Wheel C on MS (a,c,e) with vertical loads of 80N, 130N,150N,190N (light to dark) and MMS (b,d,f) with vertical loads of 80N, 110N,150N,190N (light to dark). MS experiments were performed ten times and boxplots present the average reading and one standard deviation. The relevance of these results lies in the fact that terrain characterization for the MS and MMS was not performed according to standard procedures utilized by RFT. Hence, this analysis shows the full potential of RFT applicability to generic granular materials, wheel geometry, and loading conditions.}
\label{fig:Wheel_C_MS_MMS}
\end{figure}

\begin{table}[!h]
\centering
\caption{Performance metrics for the RFT predictions of wheel C on MS and MMS. The mean absolute error $\Delta$ has dimension of [N] for drawbar, [Nm] for torque, and [mm] for sinkage. Coefficient of correlation $R$ and coefficient of variation $CV$ are unitless.}
\centering
\vspace{0.5em}

\begin{tabular}{c|cc|cc|cc|cc}
	\multicolumn{9}{c}{MS} \\ \hline
&	\multicolumn{2}{c}{80 N} &	\multicolumn{2}{c}{130 N}
&	\multicolumn{2}{c}{150 N} &	\multicolumn{2}{c}{190 N}\\ \hline
&			RFT  & MPM 	&	RFT  & MPM 	&	RFT & MPM 	&	RFT & MPM 	 \\ \hline
&	\multicolumn{8}{c}{Drawbar} \\ 
$\Delta$ &	5.55 & 6.42	& 	6.16 & 9.74	&	6.33 & 11.78& 	7.11 & 14.21 \\	
$R$ 	 &	1.00 & 0.96	& 	0.99 & 0.97	&	0.99 & 0.97 & 	0.98 & 0.96 \\	
$CV$ 	 &	0.09 & 0.10	& 	0.08 & 0.11	&	0.07 & 0.12 & 	0.07 & 0.12 \\	
\hline

&	\multicolumn{8}{c}{Torque} \\ 
$\Delta$ &	1.24 & 0.93	& 	1.79 & 1.17	&	2.11 & 1.39 & 	2.82 & 1.38 \\	
$R$ 	 &	0.99 & 0.97	& 	1.00 & 0.97	&	1.00 & 0.97 & 	1.00 & 0.98 \\	
$CV$ 	 &	0.39 & 0.29	& 	0.36 & 0.26	&	0.36 & 0.26 & 	0.38 & 0.20 \\	
\hline

&	\multicolumn{8}{c}{Sinkage} \\ 
$\Delta$ &	2.74 & 2.70	& 	3.62 & 3.45	&	3.67 & 2.41 & 	2.81 & 1.69 \\	
$R$ 	 &	0.86 & 0.93	& 	0.72 & 0.86	&	0.85 & 0.97 & 	0.92 & 0.95 \\	
$CV$ 	 &	0.28 & 0.30	& 	0.28 & 0.23	&	0.26 & 0.14 & 	0.18 & 0.09 \\	
\hline
	\multicolumn{8}{c}{} \\
	\multicolumn{9}{c}{MMS} \\ \hline
&	\multicolumn{2}{c}{80 N} &	\multicolumn{2}{c}{110 N}
&	\multicolumn{2}{c}{150 N} &	\multicolumn{2}{c}{190 N}\\ \hline
&			RFT  & MPM 	&	RFT  & MPM 	&	RFT & MPM 	&	RFT & MPM 	 \\ \hline
&	\multicolumn{8}{c}{Drawbar} \\ 
$\Delta$ &	11.42& 12.11& 	13.03 &16.26&	14.01& 21.91& 	15.11& 23.11 \\	
$R$ 	 &	0.97 & 0.93	& 	0.97 & 0.93	&	0.97 & 0.95 & 	0.95 & 0.93 \\	
$CV$ 	 &	0.15 & 0.17	& 	0.14 & 0.18	&	0.13 & 0.21 & 	0.14 & 0.21 \\	
\hline

&	\multicolumn{8}{c}{Torque} \\ 
$\Delta$ &	0.93 & 1.43	& 	1.42 & 1.98	&	1.76 & 2.20 & 	2.22 & 2.62 \\	
$R$ 	 &	0.98 & 0.94	& 	0.99 & 0.94	&	1.00 & 0.94 & 	0.99 & 0.95 \\	
$CV$ 	 &	0.31 & 0.56	& 	0.36 & 0.46	&	0.33 & 0.41 & 	0.32 & 0.40 \\	
\hline

&	\multicolumn{8}{c}{Sinkage} \\ 
$\Delta$ &	5.75 & 5.33	& 	5.85 & 3.52	&	7.47 & 4.86 & 	18.41 & 10.80 \\	
$R$ 	 &	0.50 & 0.44	& 	0.70 & 0.88	&	0.85 & 0.97 & 	0.93 & 0.90 \\	
$CV$ 	 &	1.37 & 1.45	& 	0.59 & 0.45	&	0.55 & 0.31 & 	0.56 & 0.33 \\	

\end{tabular}
\label{tab:MS_MMS_C_statistics}
\end{table}

We performed a set of experiments with wheel C on MS and MMS, for a wide range of vertical loads ranging between 80 N and 190 N. This data set was collected on a testbed which did not allow for a specified packing state. However, terrain was carefully prepared between tests in order to achieve repeatable consistent loosely packed conditions. The relevance of using RFT for these experiments lies in the fact that terrain characterization for the MS and MMS was not performed according to the standard procedures utilized by RFT for poppy seeds. The force response surfaces for these materials were obtained using scaling of similar response surfaces for PS using corresponding scaling parameters presented in Table ~\ref{tab:mech_properties}. Hence, this analysis shows the full potential of RFT applicability to generic granular materials. For this analysis, TM modeling was not done but continuum analysis (MPM) for these experiments was done in a similar fashion as before.

Figure \ref{fig:Wheel_C_MS_MMS} presents the results for four vertical loads (80 N, 110/130 N, 150 N, 190N) for wheel C traveling on MS (a,c,e) and MMS (b,d,f). For the MS sand, both the RFT and MPM underestimate (in absolute value) drawbar pull, while they both underestimate torque for positive slip only. Sinkage predictions are accurate for the whole slip range with mean absolute error in the range of 4-5 mm. For the MMS sand, similar trends are observed with less accuracy and larger absolute errors at high positive slips.

Table \ref{tab:MS_MMS_C_statistics} presents the values of mean absolute error, coefficient of correlation, and coefficient of variation for the RFT and MPM model. Regardless of the quantity under consideration, RFT performs better at lower vertical loads in terms of mean absolute error than MPM. This is also partially true for sinkage. The coefficient of correlation is above 0.85 in all cases except one. In general, coefficient of variation decreases with increasing load for both the models. Increased variability in the sinkage measurements comes from the uncertainty in controlling the terrain's free surface level and flatness. With increasing load, sinkage increases, which then leads to decreased relative errors. The performance of MPM is comparable to RFT in most cases and is observed to be better in a few cases (based on CV and R data).

\section{Comparison between RFT and TM}
Although the MS and MMS sands were characterized following best practices for TM models, results obtained with the TM model remain inaccurate when using the shear modulus obtained from direct shear tests. As discussed in \cite{ROB:ROB21483}, the shear modulus calculated from direct shear tests is in the order of tenths of millimeters, creating unrealistically high drawbar and torque predictions. However, even if the shear modulus is treated as a tuning parameter, TM predictions generally remain less accurate than RFT. For this analysis, predictions for the TM models are provided with the measured shear displacement modulus and a modulus of 0.015 m (the corresponding results are labeled TM*). The discussion below is based on the results obtained with the larger shear modulus.

\begin{figure}[htbt]
\centering
\subfigure[]{
\includegraphics[trim = 1mm 1mm 13mm 1mm, clip, width = 0.3\textwidth]{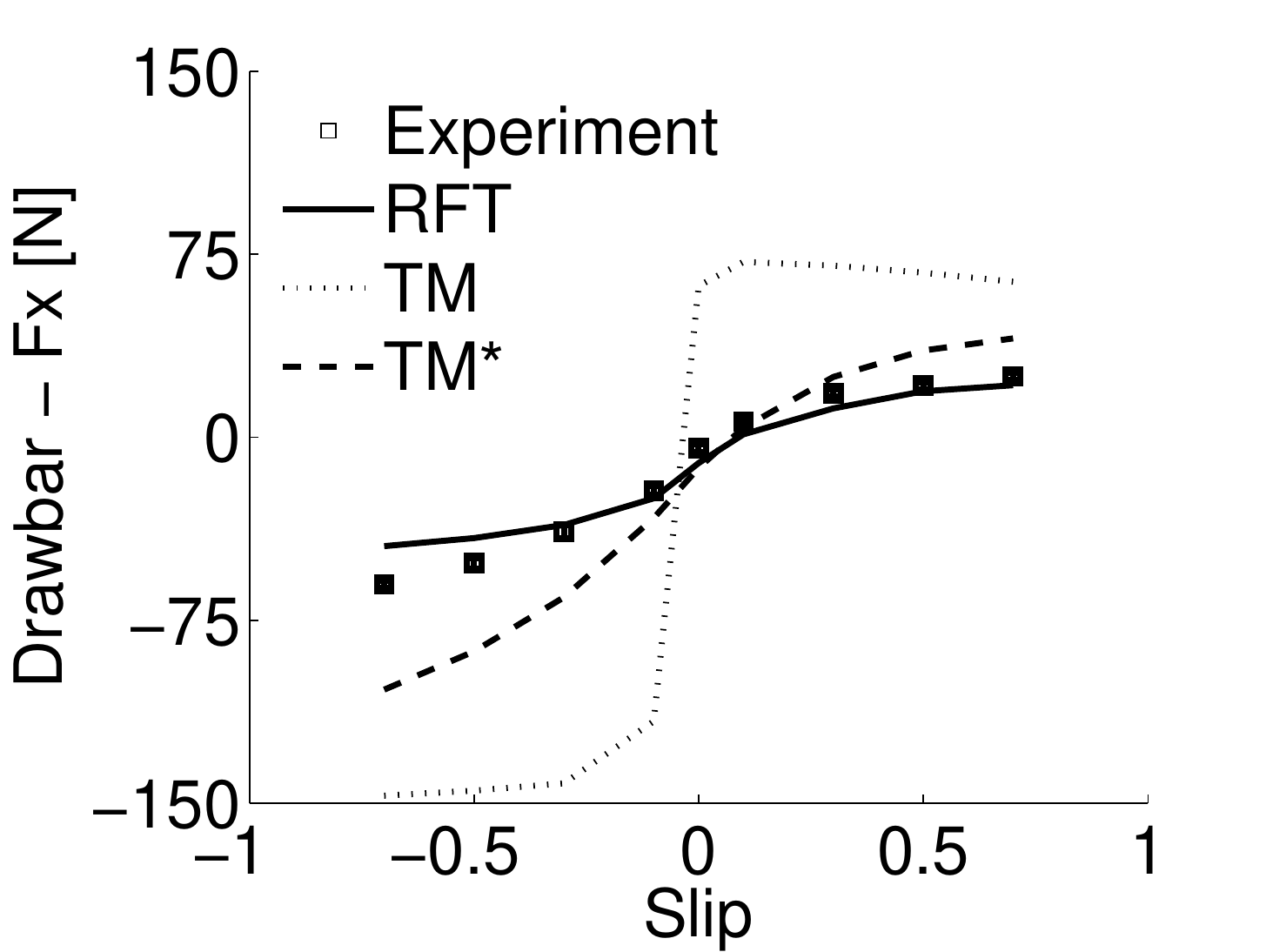}
\label{fig:DP_WheelC_130_MS}
}
\subfigure[]{
\includegraphics[trim = 1mm 1mm 13mm 1mm, clip, width = 0.3\textwidth]{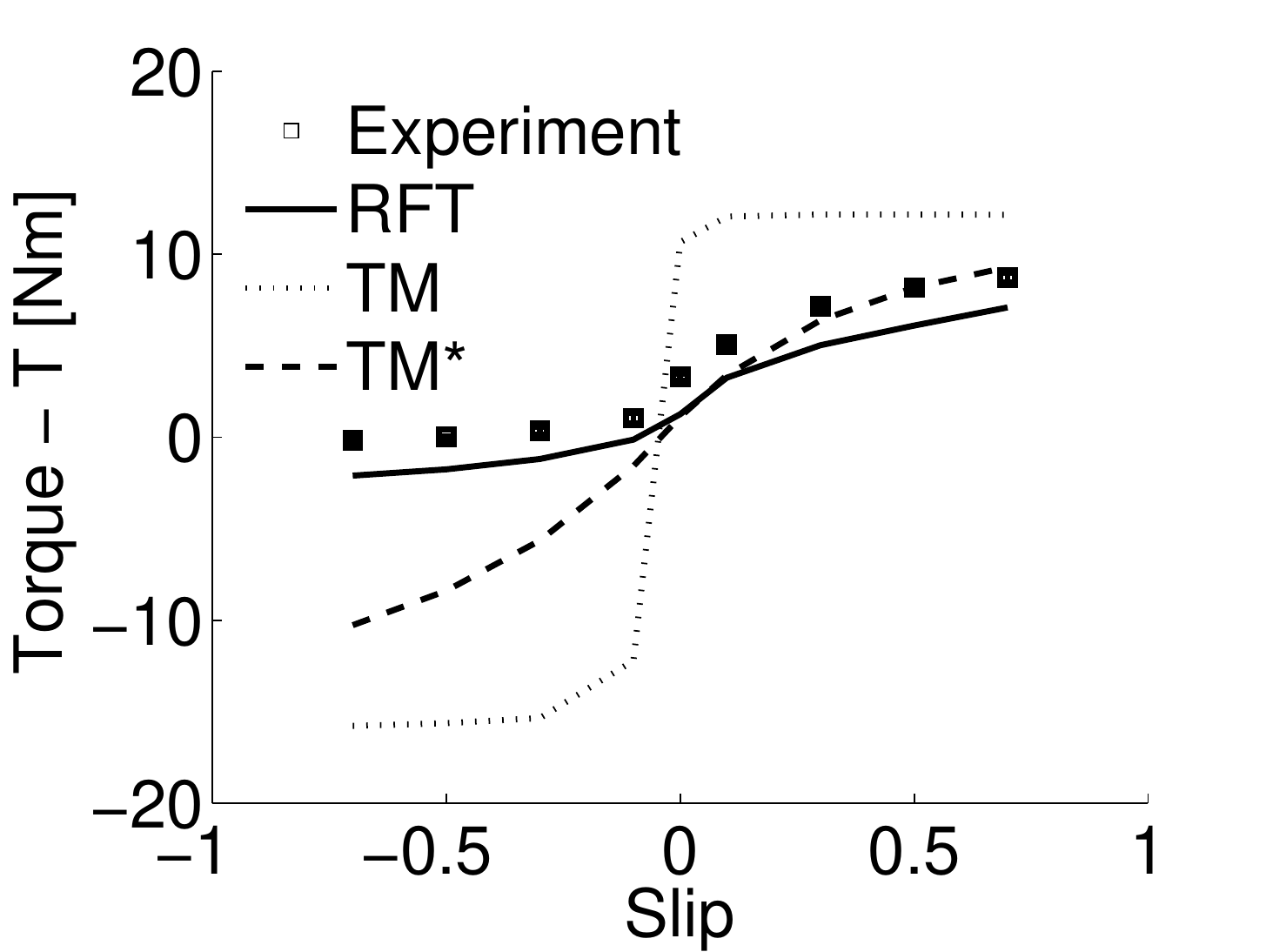}
\label{fig:T_WheelC_130_MS}
}
\subfigure[]{
\includegraphics[trim = 1mm 1mm 13mm 1mm, clip, width = 0.3\textwidth]{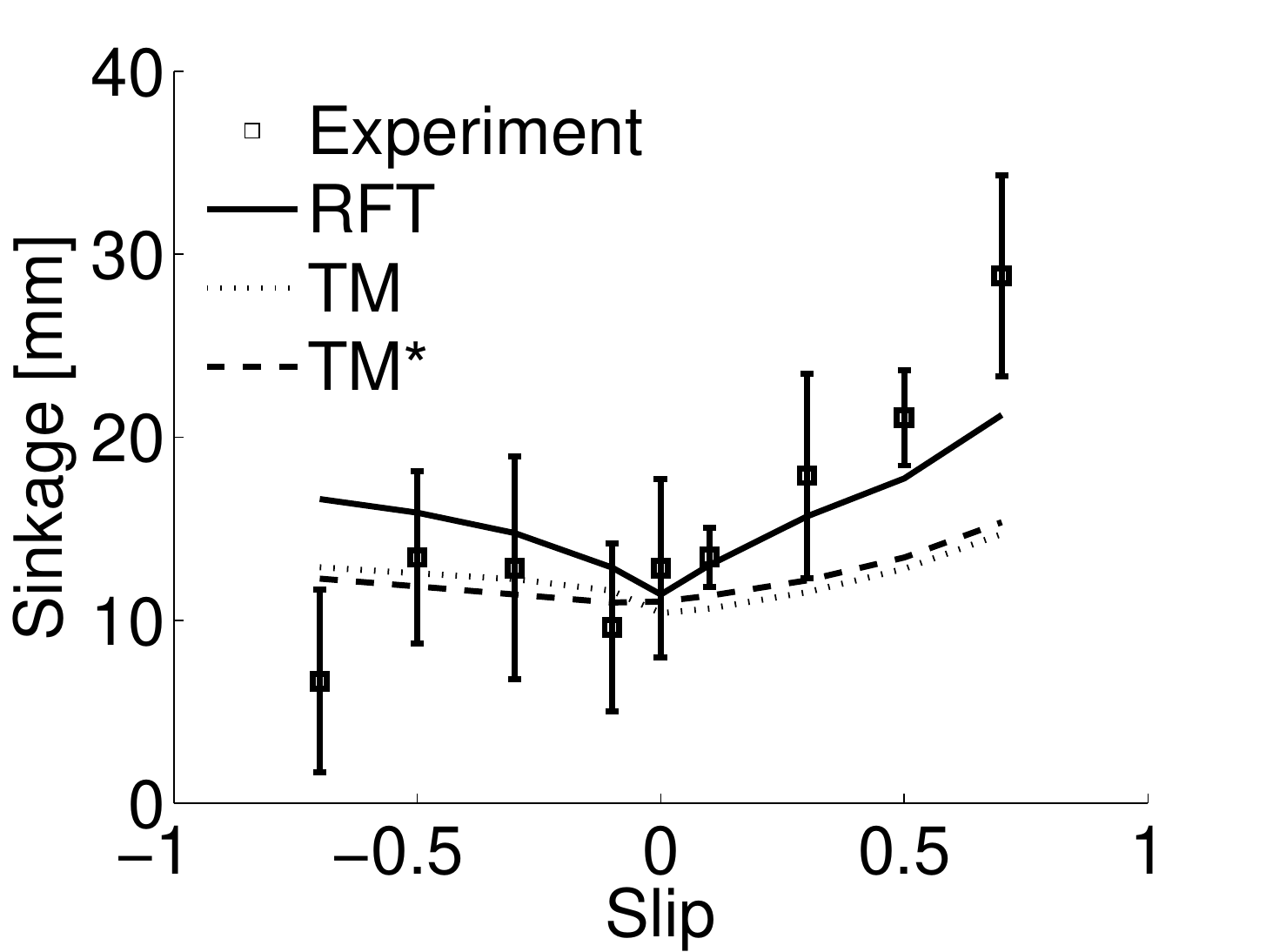}
\label{fig:Z_WheelC_130_MS}
}
\caption{Comparison of RFT and TM model for wheel C on MS. Experiments were repeated ten times and boxplots present the average reading and one standard deviation. Nominal vertical load is 130 N. Resistive force theory (RFT) and terramechanics (TM) approaches produce similar predictions for drawbar and torque at positive slip, while RFT outperforms the TM model when sinkage is considered.}
\label{fig:WheelC_130_MS}
\end{figure}

Results presented in Figure \ref{fig:WheelC_130_MS} show performance for wheel C under 130 N of vertical load while traveling on MS. Table \ref{tab:MS_C_statistics} presents the values of mean absolute error, coefficient of correlation, and coefficient of variation for the RFT and the TM models. When analyzing drawbar, RFT has a similar coefficient of correlation, but lower mean absolute error and coefficient of variation than the TM* model. The TM* model deviates significantly from measured data at negative slip. The situation is similar for torque. However, in this case RFT underestimates torque readings for the whole range, even if it maintains a high coefficient of correlation at 0.99. 

The analysis is more intricate when sinkage is considered. Qualitatively, the TM* model accurately describes the data at low negative slip, while RFT predictions are closer at positive slip levels. As a result, the TM* and RFT models have similar metrics with a mean absolute error close to 4 mm, a coefficient of variation below 0.4, and a coefficient of correlation above 0.7 for both.

TM* model performance for drawbar and torque predictions is similar to RFT when only positive slip is considered. This is relevant because for design and evaluation purposes, performance between 10\% and 30\% slip are typically used as indicators. However, as shown by the wheel--terrain configurations previously discussed, sinkage predictions were inaccurate when the TM* model was used.

\begin{table}[!h]
\centering
\caption{Comparison of resistive force theory (RFT) and terramechanics (TM) models predictions for wheel C on MS under 130 N nominal load. Columns labeled TM* refer to the TM model with shear displacement modulus $K$ = 0.015 m. The mean absolute error $\Delta$ has a dimension of [N] for drawbar, [Nm] for torque, and [mm] for sinkage. Coefficients of correlation $R$ and coefficients of variation $CV$ are unitless.}
\centering
\vspace{0.5em}

\begin{tabular}{c|ccc|ccc|ccc}
&	\multicolumn{3}{c}{Drawbar} &	\multicolumn{3}{c}{Torque} &	\multicolumn{3}{c}{Sinkage} \\ \hline
&	RFT &	TM & TM* &	RFT &	TM & TM* &	RFT &	TM & TM* \\ \hline
$\Delta$	&	6.16	&	71.84	& 18.30	&	1.79	&	9.66	&	3.59	&	3.62	&	4.84	&	4.52	\\	
$R$			&	0.99	&	0.93	& 1.00	&	1.00	&	0.91	&	0.97	&	0.72	&	0.57	&	0.85	\\
$CV$		&	0.08	&	0.76	& 0.23	&	0.36	&	2.14	&	0.98	&	0.28	&	0.39	&	0.36	\\

\end{tabular}
\label{tab:MS_C_statistics}
\end{table}

\section{Conclusion}

In this paper we analyzed the performance of resistive force theory and continuum plasticity modeling for the problem of predicting rigid wheel--dry granular media interaction. Upon comparison of experimental data for three differently shaped rigid wheels under forced---slip and variable load conditions, we concluded that though RFT was originally developed for studying legged locomotion on granular media, it can also be used as a qualitatively and sometimes quantitatively accurate model for the locomotion of rigid wheels on granular materials. The current work also establishes plasticity-based continuum modeling using an MPM implementation as a suitable candidate for  predicting wheel performance. MPM studies give complete flow and stress fields which gives deeper insight about the system, which is of vital importance for improving our understanding of locomotion processes.  
     
      Quantitative comparison was done by comparing experimental drawbar, torque, and sinkage data with model predictions and with predictions of a more traditional terramechanics model. The torque and sinkage predicted by RFT as well as the continuum model was found to have lower mean absolute errors and coefficients of variation than TM for motion on loosely/closely packed poppy seeds. MPM was found to have higher accuracy than RFT in predicting torque as well as sinkage values in general (except for wheel A on PS). Drawbar values calculated with RFT, MPM, as well as TM models are close to the experimental measurement (within 25\%), with TM having better accuracy than the other two methods and RFT having better accuracy than MPM. Considering the empirical nature of the TM model, it should be possible to obtain better predictions by performing an ad-hoc characterization of the PS simulant. In fact, by performing pressure-sinkage experiments using a plate with an area comparable with the contact patch of the wheels under consideration, it should be possible to obtain pressure-sinkage parameters that result in more accurate TM model predictions. 

By extrapolating the response surfaces from poppy seeds, we used RFT to model forced-slip experiments of rigid wheels on two natural sands (MS and MMS). RFT predictions were in qualitative agreement with experiments, exhibiting overall better performance when compared to the TM model. However, quantitative disagreement between models and experiments across the whole slip range remains. For example, drawbar in RFT is generally overestimated (in absolute value) at large slip ratios ($> 0.4$ or $< -0.4$), and underestimated at low positive slip ratios (between 0 and 0.2). While RFT calculated sinkage values are accurate for MS for the whole slip range under all wheel loadings, for MMS the predictions show significant deviations, particularly under large loadings. For MS and MMS, it remains to be seen how the accuracy of RFT predictions is affected if the response surfaces are directly obtained (in the current study response surfaces were extrapolated from downward intrusion tests).

It should be noted that the MPM deviations from experimental data observed in this study are in part attributable to the fact that the MPM implementation here was done assuming the granular motion to be plane-strain in all the test cases. This assumption can only be fully justified in low sinkage cases where the out-of-plane width of the contact patch of the wheel interface is much larger than its in--plane width. A fully 3--dimensional model could  help eliminate the issue. Thus, while this study focused on quasi-static forced--slip wheel behaviors, future studies are planned to experimentally and computationally explore the angular velocity driven, free locomotion of rigid wheels (in 3D) at wider ranges of angular velocities on poppy seeds as well as other simulants to explore high speed locomotion dynamics as well as the capability of these methods in modeling different scenarios.  We also plan to use advanced experimental methods, as in \cite{senatore2012investigation}, to allow us to validate MPM predictions for the sub-wheel flow field and traction distributions in different cases.  The ability to model interactions with non-rigid wheels is another key direction worth testing; a potential variant of the numerical method in ~\cite{recuero2017high} could permit the modeling of deformable tires to be coupled with the MPM soil treatment.

\section{Acknowledgements}
SA and KK acknowledge support from  Army Research Office (ARO) grants W911NF1510196 and W911NF1810118 and support from the U.S. Army Tank Automotive Research, Development and Engineering Center (TARDEC).  KI acknowledges support from TARDEC and ARO grant W911-NF1310063. DG acknowledges support from ARO and TARDEC.

\section{References}
\bibliography{rftpaperbib}

\bibliographystyle{ieeetr}
\end{document}